\documentstyle[psfig,general_cite,longtable,graphicx]{mn}
\newif\ifAMStwofonts

\def\arcm{\hbox{$^\prime$}}
\def\etal{{\rm et al.}\thinspace}
\def\eg{{\it e.g.\ }}

\def\ie{{\it i.e.\ }}

\def\deg{\hbox{$^\circ$}}
\def\lxlb{L$_X$/L$_B$~}
\def\lxlbtwo{L$_X$:L$_B$~}

\def\LBsol{L$_{B\odot}$~}
\def\Ldscr{L$_{dscr}$~}
\def\spose#1{\hbox to 0pt{#1\hss}}
\def\gtsim{$\mathrel{\spose{\lower 3pt\hbox{$\sim$}}
        \raise 2.0pt\hbox{$>$}}$\thinspace}
\ifoldfss
  \ifCUPmtlplainloaded \else
    \NewTextAlphabet{textbfit} {cmbxti10} {}
    \NewTextAlphabet{textbfss} {cmssbx10} {}
    \NewMathAlphabet{mathbfit} {cmbxti10} {} 
    \NewMathAlphabet{mathbfss} {cmssbx10} {} 
  \fi
  \ifAMStwofonts
    \ifCUPmtlplainloaded \else
      \NewSymbolFont{upmath} {eurm10}
      \NewSymbolFont{AMSa} {msam10}
      \NewMathSymbol{\upi}     {0}{upmath}{19}
      \NewMathSymbol{\umu}     {0}{upmath}{16}
      \NewMathSymbol{\upartial}{0}{upmath}{40}
      \NewMathSymbol{\leqslant}{3}{AMSa}{36}
      \NewMathSymbol{\geqslant}{3}{AMSa}{3E}

      \let\leq=\leqslant 
      \let\geq=\geqslant 
    \fi
  \fi
\fi 

\ifnfssone
  \newmathalphabet{\mathit}
  \addtoversion{normal}{\mathit}{cmr}{m}{it}
  \addtoversion{bold}{\mathit}{cmr}{bx}{it}
  \newmathalphabet{\mathbfit} 
  \addtoversion{normal}{\mathbfit}{cmr}{bx}{it}
  \addtoversion{bold}{\mathbfit}{cmr}{bx}{it}
  \newmathalphabet{\mathbfss} 
  \addtoversion{normal}{\mathbfss}{cmss}{bx}{n}
  \addtoversion{bold}{\mathbfss}{cmss}{bx}{n}
  \ifAMStwofonts
    \ifCUPmtlplainloaded \else
      \UseAMStwoboldmath
      \makeatletter
      \new@mathgroup\upmath@group
      \define@mathgroup\mv@normal\upmath@group{eur}{m}{n}
      \define@mathgroup\mv@bold\upmath@group{eur}{b}{n}
      \edef\UPM{\hexnumber\upmath@group}
      \new@mathgroup\amsa@group
      \define@mathgroup\mv@normal\amsa@group{msa}{m}{n}
      \define@mathgroup\mv@bold\amsa@group{msa}{m}{n}
      \edef\AMSa{\hexnumber\amsa@group}
      \makeatother
      \mathchardef\upi="0\UPM19
      \mathchardef\umu="0\UPM16
      \mathchardef\upartial="0\UPM40
      \mathchardef\leqslant="3\AMSa36
      \mathchardef\geqslant="3\AMSa3E

      \let\leq=\leqslant 
      \let\geq=\geqslant 
    \fi
  \fi
\fi 

\ifnfsstwo
  \DeclareMathAlphabet{\mathbfit}{OT1}{cmr}{bx}{it}
  \SetMathAlphabet\mathbfit{bold}{OT1}{cmr}{bx}{it}
  \DeclareMathAlphabet{\mathbfss}{OT1}{cmss}{bx}{n}
  \SetMathAlphabet\mathbfss{bold}{OT1}{cmss}{bx}{n}
  \ifAMStwofonts
    \ifCUPmtlplainloaded \else
      \DeclareSymbolFont{UPM}{U}{eur}{m}{n}
      \SetSymbolFont{UPM}{bold}{U}{eur}{b}{n}
      \DeclareSymbolFont{AMSa}{U}{msa}{m}{n}
      \DeclareMathSymbol{\upi}{0}{UPM}{"19}
      \DeclareMathSymbol{\umu}{0}{UPM}{"16}
      \DeclareMathSymbol{\upartial}{0}{UPM}{"40}
      \DeclareMathSymbol{\leqslant}{3}{AMSa}{"36}
      \DeclareMathSymbol{\geqslant}{3}{AMSa}{"3E}

      \let\leq=\leqslant 
      \let\geq=\geqslant 
    \fi
  \fi
\fi 

\ifCUPmtlplainloaded \else
  \ifAMStwofonts \else 
    \def\upi{\pi}
    \def\umu{\mu}
    \def\upartial{\partial}
  \fi
\fi
\def\etal{{\it et al. }}

\begin{document}

\title[
A Catalogue and Analysis of 
X--ray luminosities of Early--type galaxies 
]
{
A Catalogue and Analysis of 
X--ray luminosities of Early--type galaxies
}
\author[
Ewan O'Sullivan et al. 
]
{
Ewan O'Sullivan$^{1}$, Duncan A. Forbes$^{1,2}$,
Trevor J. Ponman$^{1}$\\
$^1$School of Physics and Astronomy, 
University of Birmingham, Edgbaston, Birmingham B15 2TT \\
(E-mail: ejos@star.sr.bham.ac.uk)\\
$^2$Astrophysics \& Supercomputing, Swinburne University,
Hawthorn VIC 3122, Australia\\
\\
}

\pagerange{\pageref{firstpage}--\pageref{lastpage}}
\def\LaTeX{L\kern-.36em\raise.3ex\hbox{a}\kern-.15em
    T\kern-.1667em\lower.7ex\hbox{E}\kern-.125emX}

\newtheorem{theorem}{Theorem}[section]

\label{firstpage}

\maketitle

\begin{abstract}

We present a catalogue of X--ray luminosities for 401 early--type galaxies,
of which 136 are based on newly analysed {\it ROSAT} PSPC pointed
observations. The remaining luminosities are taken from the literature and
converted to a common energy band, spectral model and distance scale. Using 
this sample we fit the \lxlbtwo relation for early--type galaxies and find
a best fit slope for the catalogue of $\sim$2.2. We demonstrate the
influence of group--dominant galaxies on the fit and present evidence that
the relation is not well modeled by a single powerlaw fit. We also derive
estimates of the contribution to galaxy X--ray luminosities from 
discrete sources and conclude that they provide L$_{dscr}$/L$_B\simeq$29.5
erg s$^{-1}$ L$_{B\odot}^{-1}$. We compare this result to luminosities from
our catalogue. Lastly, we examine the influence of environment on galaxy
X--ray luminosity and on the form of the \lxlbtwo relation. We conclude
that although environment undoubtedly affects the X-ray properties of
individual galaxies, particularly those in the centres of groups and
clusters, it does not change the nature of whole populations. 

\end{abstract}


\section{Introduction}

One of the most surprising results from the {\it Einstein} observatory
(launched in 1978) was the discovery of diffuse X--ray emission from
early--type 
galaxies. Since then, many X--ray 
studies of galaxies have been published, ranging 
between detailed analyses of individual objects and large catalogues
designed to shed light on their global
properties. \scite{fabbianokimtrinchieri92} (FKT) produced one of the most
extensive catalogues using {\it Einstein} data, 
observing 148 early--type galaxies and
examining (among other things) the \lxlbtwo relation for these objects. 
Other works in a similar vein include 
 those of \scite{Bursteinetal97}, a somewhat larger catalogue of
 {\it Einstein} data, \scite{daviswhite96} and
\scite{brownbreg98} which use smaller samples based on {\it ROSAT} PSPC pointed
observations, \scite{Beuingetal99} based on the {\it ROSAT} All--Sky Survey, and
\scite{Matsushita00b} using {\it ROSAT} pointed data. 

The largest of these catalogues, that of Beuing \etal, contains almost 300
galaxies, but most of these have exposure times of only a few hundred
seconds. Catalogues based on pointed data have much longer
exposures, but lack the coverage to be truly representative of the general
population of early--type galaxies. The problem is exacerbated by the fact
that most small and medium sized samples focus on the brightest objects,
and pass over the fainter and less well studied galaxies. It can also be
difficult to compare between catalogues, as each employs its own analysis
procedure and presents results in its own particular format. For example,
we have not used data from the sample of \scite{Bursteinetal97} because the
method used to convert count rates to fluxes is not based on a single
spectral model, making it more difficult to correct luminosities
from this catalogue to our own model and waveband. 

Our intention in this paper is to provide a large general catalogue of X--ray
luminosities for early--type galaxies. We have therefore calculated new
X--ray luminosities for 136 galaxies, based on {\it ROSAT} PSPC data, and
added a further 265 luminosities from previously published catalogues. All
of the X--ray luminosities have been converted to a common format based on a
reliable distance scale (assuming H$_0$ = 75 km s$^{-1}$
Mpc$^{-1}$) and correcting for differences in spectral fitting techniques
and waveband. We use the resulting catalogue to study the
X--ray properties of early--type galaxies, focusing in particular on the
\lxlbtwo relation and on the influence of environment. 

In Section~\ref{samplesel} we give details of our sample, and discuss our
X--ray analysis of {\it ROSAT} data in Section~\ref{Specfit}. Section~\ref{catconv}
covers the methods used to add data from the literature to our own results, 
and Section~\ref{surv} briefly discusses the survival analysis techniques
used to fit lines to censored data. In Section~\ref{Res2} we report the
results of our line fits to the \lxlbtwo relation, as well as giving an
estimate of the contribution of discrete sources and examining the
influence of galaxy environment. Section~\ref{discuss} is a discussion of
some of our results and the conclusions we draw from them. Throughout the
paper we normalise L$_B$ using the
solar luminosity in the B band, L$_{B\odot}$ = 5.2 $\times$ 10$^{32}$
erg s$^{-1}$. 

\section{Sample selection} 
\label{samplesel}

Our sample of early--type galaxies was selected from the Lyon--Meudon
Extragalactic Data Archive (LEDA). This catalogue contains information on
$\sim$100,000 galaxies, of which $\sim$40,000 have redshift and
morphological data. Galaxies were selected using the following criteria:

\begin{itemize}
\item Morphological Type T $< -1.5$ ($\ie$ E, E--S0 and S0 galaxies)
\item Virgo corrected recession velocity V $\leq$ 9,000 km s$^{-1}$
\item Apparent Magnitude B$_T \leq$ 13.5
\end{itemize}

The redshift and apparent magnitude restrictions were chosen in order to
minimise the effects of incompleteness on our sample. The LEDA catalogue is 
known to be 90\% complete at B$_T$ = 14.5 (\pcite{Amendola97}), so our
selection should be close to statistical completeness. 
The selection process produced $\sim$700 objects. We then cross--correlated
this list with a list of public {\it ROSAT PSPC} pointings. Only pointings
within 30$\arcm$ of the target were accepted as further off axis the {\it
  PSPC} point--spread function becomes large enough to make analysis
problematic. This left us with 209 galaxies with X--ray data available.  

\section{Data Reduction and Spectral Fitting}
\label{Specfit}

Data reduction and analysis of the X--ray datasets were carried out using
the {\sc asterix} software package. Before the datasets can be used,
various sources of contamination must be removed. Possible sources include
charged particles and solar X--rays scattered into the telescope from the
Earth's atmosphere. Onboard instrumentation provides information which allows
periods of high background to be identified. The master veto counter
records the charged 
particle flux, and we have excluded all time periods during which the
master veto rate exceeds 170 count s$^{-1}$. Solar contamination
causes a significant overall increase in the X--ray event rate. To remove
this contamination we excluded all times during which the event rate
deviated from the mean by more than 2$\sigma$. This generally
removes no more than a few percent of each dataset.

After this cleaning process each dataset is binned into a 3--dimensional (x, y, 
  energy) data cube. Spectra or images can be extracted from such a cube by
collapsing it along the axes.
A model of the background is generated based on an annulus taken
 from this cube.  
We used annuli of width 0.1$\deg$, and inner radius
0.4$\deg$ where possible. In cases where this would place the annulus close 
to the source we moved the annulus, generally to r = 0.55$\deg$. To ensure that 
the background model is not biased by sources within the annulus,
an iterative process was used to remove point sources of $> 4.5\sigma$
significance. Occasionally the 
annulus lies over an area of diffuse emission, in which case we either
remove that region by hand or move the annulus to an uncontaminated region.
The only exception to this occurred in cases where the target galaxy was
surrounded by group or cluster emission. In such cases the target is
contaminated by group emission along the line of sight, increasing its
apparent luminosity. To counter this we allowed the annulus to lie over
the outer region of the group emission (unless prevented by large numbers of
point sources), thereby removing at least a part of the contamination. 
The resulting background model was then used to produce a
background-subtracted cube. Regions near the {\it PSPC} window support
structure were
removed from these images, as objects in those areas would have been
partially obscured during the observation. The cube was further corrected
for dead time and vignetting effects, and point sources were removed. 

Examination of background subtracted images allowed us to locate each
galaxy and produce a radial profile of its emission. This profile was used
to determine the radius of the region from which a spectrum was extracted,
with the cutoff radius taken at the point where the X--ray emission drops to 
the background level. We excluded 73 sources for which derived X--ray
fluxes were unreliable at this stage. Many were too
close to the support structure, or only had very poor quality data
available. Others were found to be located close to bright X--ray
sources. Galaxies in groups and clusters were only accepted if
they stood out clearly above the general cluster emission. Point sources
within the extraction region were not removed, as we considered these
likely to be part of the galaxy emission. A spectrum of this region was
then obtained by collapsing the cube along its {\it x} and {\it y} axes.

Galaxy spectra were fitted with a MEKAL hot plasma model
(\pcite{MEKAL}; \pcite{MEKAL2}). Hydrogen absorption column densities were fixed at values determined
from radio surveys (\pcite{stark92}), and temperature and metal abundance
were fixed at 1 keV and 1 solar respectively. Fitting in this way
provides a fairly crude measure of the bolometric X--ray flux, but allows all the
galaxy spectra to be fitted by the same model, regardless of the quality of 
the data available. 

Our choice of temperature and metallicity for these fits was influenced by
our intention to combine our results with those of other studies. The
catalogues of \scite{Beuingetal99} and
\scite{fabbianokimtrinchieri92} both 
assume these values in their fits to early-type objects, although they use
a Raymond \& Smith (\pcite{R&S}) plasma model rather than the more accurate
MEKAL model. There is a strong body of evidence showing that these 
parameters are representative of the majority of early-type
galaxies. The recent study by \scite{Matsushita00}, uses high quality {\it
  ASCA} observations to examine the gas 
metallicity in a number of X--ray luminous
early-type galaxies. Taking into account probable errors in the modeling
of the Fe--L spectral region, average metal abundances are found to be
solar to within a factor of two, regardless of the plasma code
used. Measured temperatures of early-type galaxies usually range between
0.2 and 1.3 keV (\eg \pcite{Matsushita00b};
\pcite{daviswhite96}), but finding an accurate average is hampered by the
lack of high quality data. 

The spectral representation we employ is clearly over--simplistic given that 
these objects are probably better fit by two component 
models (\pcite{Matsumotoetal97}). While such multi--temperature 
models should give more
accurate measurements of halo gas temperatures, they require higher quality 
data, and have been used to date on only relatively small samples of bright
galaxies. 
Single temperature 1 keV models are most likely to be poor
descriptions of 
X-ray faint galaxies, which are expected to 
be dominated by emission from X-ray binaries (\pcite{Matsumotoetal97}). In
these galaxies, emission is likely to be better represented by a high
temperature bremsstrahlung model. If we assume that our lowest luminosity
galaxies are
actually better described by a 10 keV bremsstrahlung model, we find that we 
will have underestimated their bolometric luminosity by a factor of $\sim$2. 

In total we fitted luminosities for 136 early--type galaxies, of which only 
15 are upper limits. These form the core of our catalogue.

\section{A Master Catalogue}
\label{catconv}
Comparison of our new 
data with previously published catalogues was hampered by the
different basic parameters used in these catalogues. The three
catalogues we examined are those of \scite{Beuingetal99},
\scite{fabbianokimtrinchieri92} and \scite{robertshogg}. These
use a range of
models to fit the data, different wavebands, 
distances and blue luminosities. We have
corrected for these differences by converting the catalogues to a
common set of values, as used for our own results.

Where possible, we take our distances from \scite{pands96}. These are computed 
using the model of \scite{FaberBurstein88} which accounts for the influence of
the Great Attractor and Virgocentric flow. For galaxies not listed in
Prugniel \& Simien we used distances from LEDA, which are corrected only
for Virgocentric motion. Similarly, we have calculated L$_B$ for each
object based where possible on the B$_T$ values given in Prugniel \&
Simien. Where these are unavailable we use B$_T$, or in some cases m$_B$,
from NED. Galaxies for which we have used m$_B$ to calculate L$_B$ are
marked in the final catalogue, and in order to test their effect on our
results we compared their distribution on an \lxlbtwo graph with that of
the rest of our catalogue. We found no significant difference between the
two subsets. We therefore believe that these values provide us with a reasonably
homogeneous and accurate set of distances and luminosities on which to base
our study.

The three catalogues we wish to compare our results to each quote L$_X$ in
different wavebands. \scite{fabbianokimtrinchieri92} and
\scite{robertshogg}, working with the {\it Einstein} IPC, quote
luminosities in the 0.2--4.0 keV and 0.5--4.5 keV bands. \scite{Beuingetal99} 
choose a 0.64--2.36 keV band, as their work is based on relatively low
signal to noise ROSAT PSPC All--Sky Survey data. To allow us to compare
these with our luminosities we convert each to a pseudo--bolometric
band. The spectral models available generally have limited energy range;
for example, the Raymond \& Smith model grid available on {\sc asterix} covers
the 0.088--17.25 keV range. However, we have assumed a typical galaxy
temperature of 1 keV, as do the three other catalogues, so contributions to 
any model from outside the available range should be negligible. Using
{\sc xspec} (v11.0.1) we have tested this and find that changing the lower
bound to 10 eV has no effect increases L$_X$ by $\sim$6\%, while changing the
upper bound to 100 keV produces no measurable increase. 

We also need to correct for different spectral models. For our analysis we
have used the MEKAL model, as this is probably the most accurate
generally available.
However, both Beuing \etal and Fabbiano \etal use the 
Raymond \& Smith model, and Roberts \etal use a simple bremsstrahlung
model. Luckily, the choice of solar metallicity is common to
all. Therefore, we calculated conversion factors between 1 keV, solar
metallicity Raymond \& Smith and bremsstrahlung models in the appropriate
wavebands and our own MEKAL model in the pseudo--bolometric band. We then
apply these corrections to the catalogues, bringing their luminosities
into line with ours. The correction factors, including the effects of
plasma code and conversion to bolometric luminosities, are shown in
Table~\ref{Corrections}. 

\begin{table}
\begin{center}
\begin{tabular}{lc}
Catalogue & Correction Factor  \\
 & $\Delta$Log L$_X$ \\ 
\hline
Beuing \etal & +0.27 \\
Fabbiano \etal & +0.15 \\
Roberts \etal & +0.36 \\
\end{tabular}
\end{center}
\caption{Correction factors used to convert luminosities from Beuing \etal, 
  Fabbiano \etal and Roberts \etal into our pseudo--bolometric band and
  MEKAL model. 
\label{Corrections}  }
\end{table}

\begin{figure*}
\parbox[t]{1.0\textwidth}{\vspace{-1em}\includegraphics[width=9.5cm]{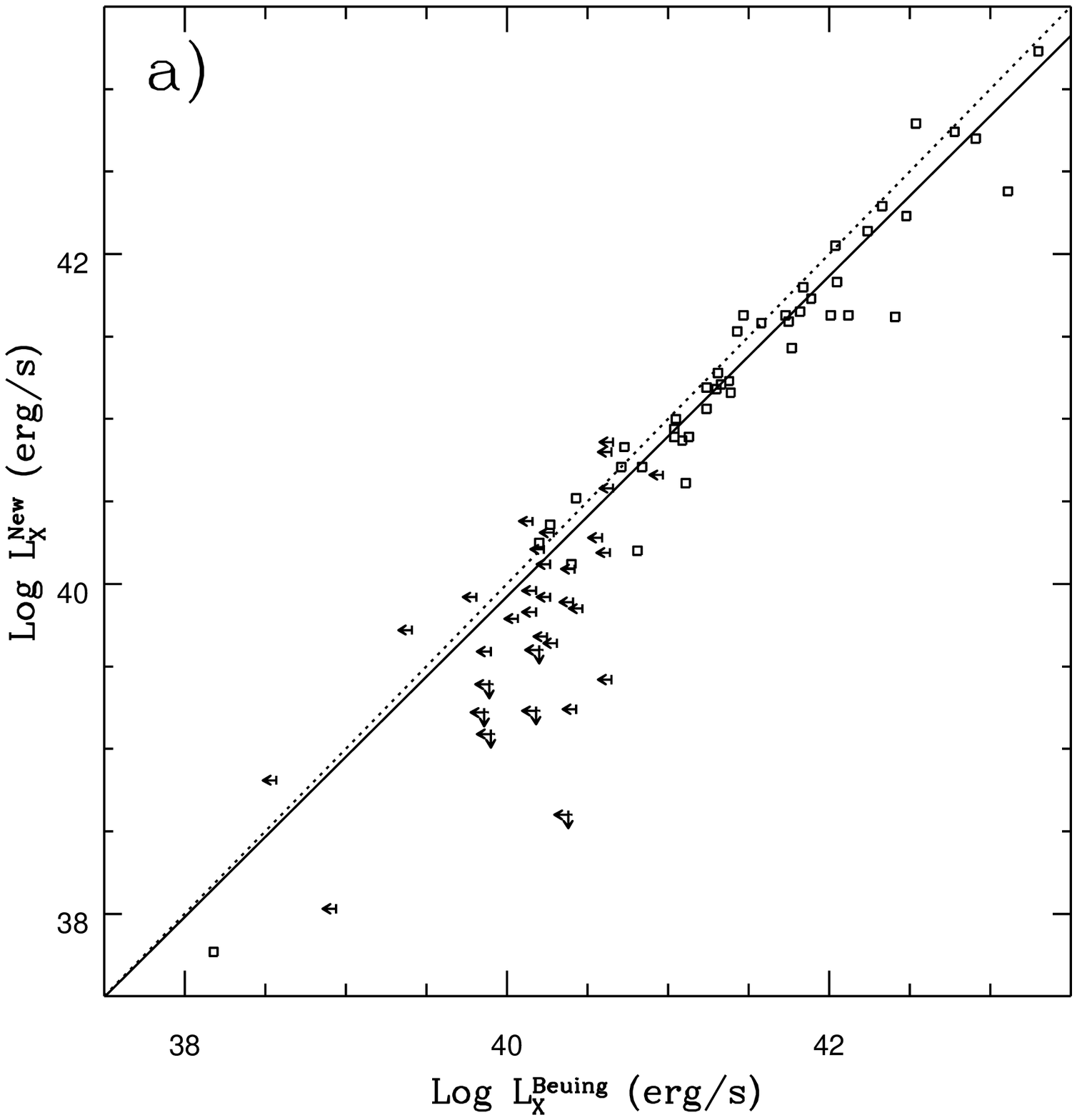}}
\parbox[t]{0.0\textwidth}{\vspace{-12.7cm}\includegraphics[width=9.5cm]{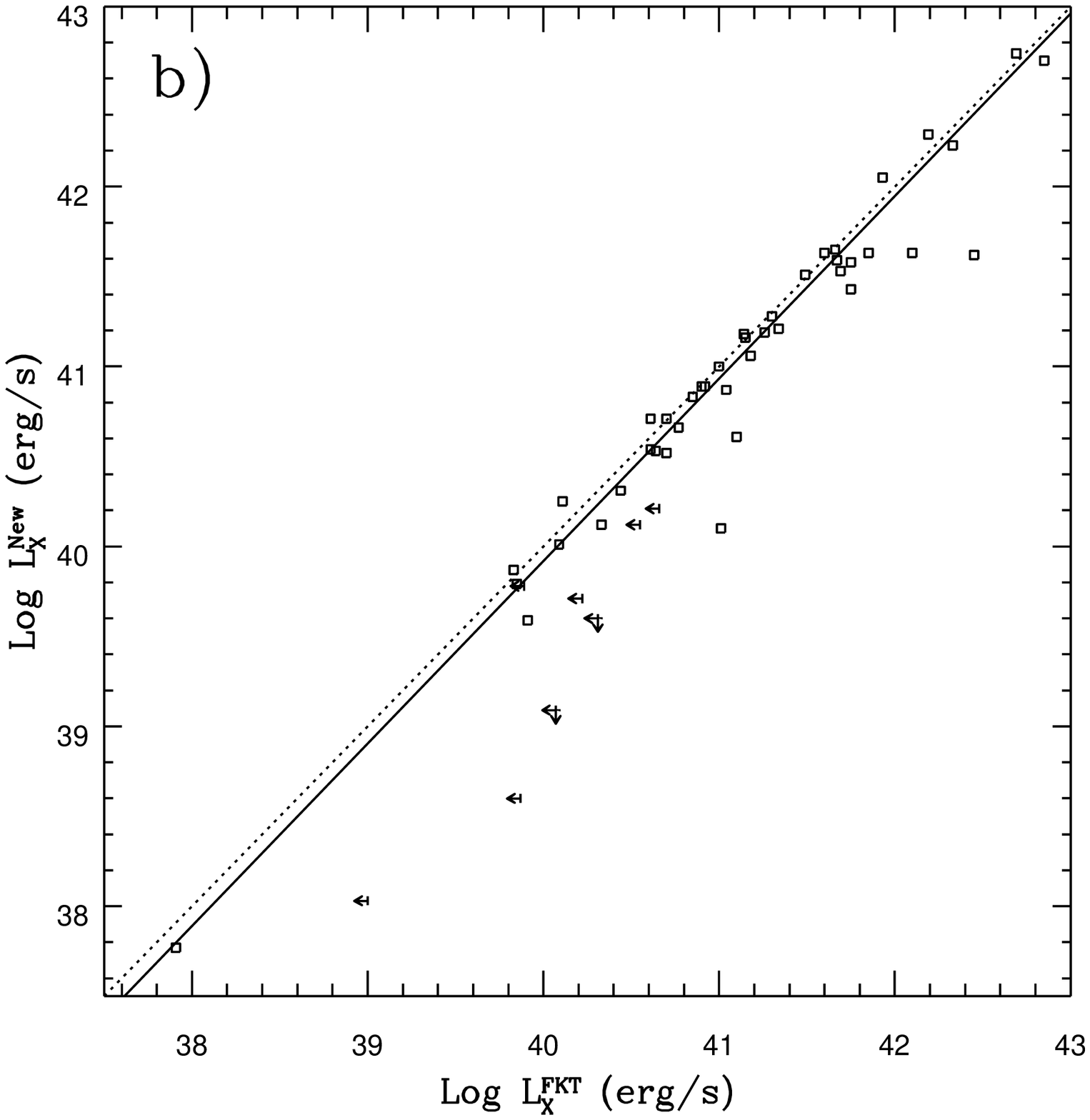}}
\parbox[t]{1.0\textwidth}{\vspace{-4cm}\includegraphics[width=9.5cm]{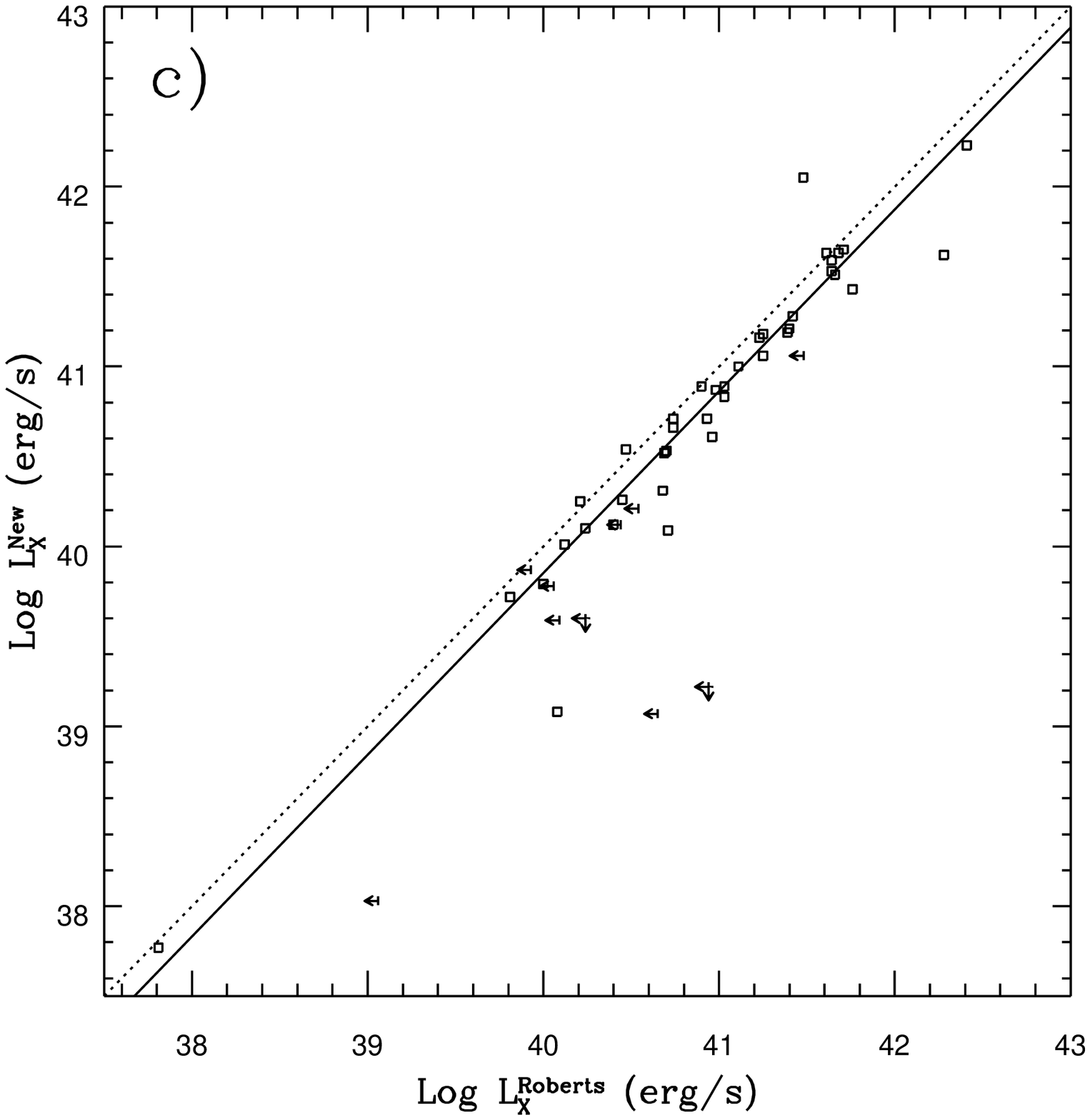}}
\vspace{-3cm}
\hspace{8.5cm}
\begin{minipage}{7cm}
\vspace{-20.7cm}
\caption{\label{Compplots}  Comparisons of corrected bolometric L$_X$
  values from our PSPC pointed data and three other catalogues; a) Beuing
  \etal (1999), b) Fabbiano, Kim \& Trinchieri (1992) and c) Roberts \etal
  (1991). Solid lines  
  are best fits to the detections, excluding aberrant points as described
  in the text. Dotted lines show 1:1 relations between each pair of
  luminosities. All points, including those excluded from the fits, are
  marked as squares (detections) or arrows (upper limits).} 
\end{minipage}
\end{figure*}

To confirm that this process acts as intended, we compare L$_X$ values for
those galaxies which are listed in more than one catalogue. Plots of these
comparisons are shown in Figure~\ref{Compplots}. 

In all three plots, a strong and fairly tight correlation is clear. In
order to establish the relation between the three catalogues and our own
points, we have fitted regression lines to the data. Galaxies actually
detected in two catalogues 
should have very similar measured luminosities. However, the differences in 
data quality between the samples imply that upper limits may not be
similar. We therefore fit the lines using detections only. We also expect a 
certain number of galaxies for which the measured luminosities
disagree. There may be cases where the lower spatial resolution of the {\it
  Einstein} IPC or the small exposure times of the RASS observations allow
confusion from nearby sources. Contamination from group or cluster emission
is also likely to be dealt with differently in the different catalogues. To 
avoid bias from such cases we therefore exclude from our fits galaxies for
which the luminosities disagree  
by more than a factor of $\sim$3. A search in NED revealed that all the
galaxies thus excluded are either AGN (such as Cen A), surrounded by group
or cluster emission (such as M86 or NGC 720) or lie near a much brighter
companion galaxy (NGC 3605). Lastly, we also exclude the local group dwarf
elliptical, M32, as it has a much lower luminosity than any of the other
galaxies, and tends to drive the fitting process.
With these galaxies excluded, we use the OLS bisector method to fit lines to
the data. The slopes and intercepts are shown in Table~\ref{complines}.

\begin{table}
\begin{center}
\begin{tabular}{lcc}
Catalogue & \multicolumn{2}{c}{Best Fit}  \\
 & Slope (Error) & Intercept (Error) \\ 
\hline
Beuing \etal & 0.971 ($\pm$0.031) & 1.057 ($\pm$1.298) \\
FKT & 1.014 ($\pm$0.028) & -0.672 ($\pm$1.160) \\
Roberts \etal & 1.011 ($\pm$0.028) & -0.593 ($\pm$1.132) \\
\end{tabular}
\end{center}
\caption{Comparison between galaxies from our PSPC pointed data
and those from other samples. 
\label{complines}  }
\end{table}

The relations between the two {\it Einstein}--based catalogues and our own
L$_X$ values both have slopes close to unity, and small intercept values. We
take this as an indication that the corrected catalogues are
comparable. In the case of the Beuing \etal luminosities we find a slope
of slightly less than unity, suggesting that their luminosities become
systematically brighter than ours at high L$_X$. We believe this to be
caused by a difference in analysis technique. Beuing \etal 
take source radii, as we do, as the radius at which the X--ray emission
drops to the background level. However, when dealing with group--dominant
galaxies they set the radius to include the group halo, whereas we attempt
to use a radius at which the galaxy emission drops to the group level. This 
means that at high L$_X$, some of their luminosities include considerably
more group emission than ours.

These relations show that our correction factors do indeed bring the
catalogues into good agreement with one another. We do however recognize
that there are likely to be factors we are unable to take into account,
such as the use of different source extraction radii, and so we apply the
relations defined in Table~\ref{complines} as a further correction factor
to the X--ray luminosities from the literature. In practice,
it should be noted that the corrections are small (generally less than
$\Delta$Log L$_X$ = 0.1) and therefore have a
minimal effect on the results presented in the rest of this paper.

\section{Statistical Analysis}
\label{surv}
Before presenting the results of our new measurements, we first
discuss the statistical techniques used to analyze the various
correlations presented in this paper. 
Throughout this study we deal 
with data which contain both upper limits and 
detections. This is unavoidable when attempting to compile a large
catalogue of galaxy X--ray luminosities. Many of the objects included only
have serendipitous pointings available, and there are a number of faint
galaxies which would require longer pointings to be detected. 

To deal with data containing upper limits, we use the survival analysis
tasks available in {\sc iraf}. Survival analysis assumes that the censoring of
the data is random -- \ie that the upper limits are unrelated to the true
values of the parameter. In more detail, the assumption is made that
for each upper limit, the distribution of detections below this
value forms a reasonable model for the probability distribution of
the true value associated with the upper limit. This assumption
would be invalidated if, for example, sensitivity limits were systematically 
related to the true fluxes from sources -- for example by observing
known faint sources for longer in an attempt to detect them.
In the case of our samples, we have upper limits
representing galaxies over the majority of the range of L$_X$, and the
detection limits are determined by exposure time, source distance, off axis 
angle and in
some cases source environment. Most of the galaxies whose X--ray
luminosities we have calculated based on {\it ROSAT} pointed data were not
the target of the pointings used. This suggests that exposure time should be
unrelated to the galaxy luminosity or distance. Similarly, luminosities
taken from the Beuing \etal sample are based on exposures whose length is
unrelated to any particular target. The situation is less clear in the case 
of the luminosities based on {\it Einstein} data, as more of these objects
are likely to have been the target of the observation. However, for the
great majority of galaxies, random censoring appears to be a fair
assumption. 

Three correlation tests are available in {\sc iraf}; the generalized
Kendall's tau, generalized Spearman's rho and Cox proportional hazard
tests. Both 
the Kendall's tau and Cox hazard tests are known to perform poorly when the 
data contains large numbers of tied values, and all three tasks function
best on large datasets (\pcite{Feigelson85}). Our samples are mainly large,
in which case we use all three tests. We quote the least favourable result
- \ie the lowest significance found. In the few cases where a
sample contains less than 30 objects we do not use the generalized
Spearman's rho test. 

To fit lines to our samples we use two of the three linear regression tasks 
available. These are the expectation and maximization (EM) algorithm and
the Buckley--James algorithm (BJ). The EM algorithm is a parametric test and
assumes that the residuals to the fitted line follow a Gaussian
distribution. The BJ method is non-parametric, using
the Kaplan--Meier estimator for the residuals to calculate the regression,
and therefore only requires the censoring distribution of the data about
the line to be random. In almost all cases we find that these two methods
agree reasonably well, and in most cases their results are nearly
identical. However, in cases where the two methods are not in close
agreement it should be noted that the BJ algorithm is probably the more
reliable of the two, as it makes no assumption about the underlying
distribution of the data. When using these tasks or the correlation tests,
we take the uncensored parameter as the independent variable and the
censored parameter as the dependent variable. The EM and BJ algorithms also
produce values for the standard deviation about the regression, giving an
estimate of the scatter in the relation.

The third linear regression task available to us is the Schmitt binning
method. This technique can deal with upper limits on both axes, which
allows allows a bisector fit to be carried out, based on fitting both y/x 
(y on x) and x/y regressions lines. 
However, the Schmitt algorithm is known to be
unreliable when used with heavily censored data (\pcite{Isobeetal86};
\pcite{schmitt85}), a result confirmed by the simulations reported
in section~\ref{fit_test}. 
We therefore do not use Schmitt binning for our analysis.

To calculate means, we use the Kaplan--Meier estimator, which produces
reliable results and error estimates except in cases where the lowest point
in the data is an upper limit. When this occurs, the mean value derived
tends to be underestimated. The estimator can also be used to effectively
fit lines of fixed slope. For example, when fitting a line of slope unity
to \lxlbtwo relations, as the mean value of the \lxlb distribution is equal
to the intercept of a slope unity line. 

\section{Tests of fitting accuracy}
\label{fit_test}
When attempting to determine the underlying relationship between 
two uncensored variables, an OLS bisector fit is
likely to be the most reliable fitting method (\pcite{Isobeetal90}). For our
censored data we have used the EM and BJ algorithms, which perform y on x
regression. An alternative to these fits is to use the Schmitt binning method
to perform y/x and x/y fits and then calculate a bisector of the two. We
have carried out fits using this method, as described in
\scite{Shapleyetal01}, on various subsamples of our
data. The slopes of these `Schmitt bisector' fits are generally somewhat
steeper than the EM and BJ fits, as might be expected. However, in many
cases the slopes are very different from those found by the other two
algorithms, and in a few cases a shallower slope is found. In order to test
how well the three algorithms measure the
underlying distribution of data, we have carried out a number of
comparisons using simulated data.

We simulate datasets by using a random number generator to produce a set of 
data points, based on a predetermined straight line relation and range of
x--values. We define a level of scatter, and points are shifted up or
down by a random distance uniformly distributed within this range. 
To censor the data, we
randomly select a number of data points and calculate a new y--axis 
value for each, corresponding to a detection threshold. If this
new value is higher than the original, the data point is declared to
be an upper limit at the new, higher value. 
The range of scatter of these detection thresholds is defined separately
from the scatter in the data values, and both have been chosen to
be comparable to that seen in our real dataset.
Datasets containing the initial ``detected'' values (i.e. without any
thresholding) are also produced, and these are fitted using a 
standard OLS bisector, as well as by the EM, BJ and
Schmitt bisector methods.

As a test of the basic fitting properties of the four techniques, we
simulated a line of slope 2, intercept 0, with x ranging between 0 and 10
and a scatter in the points of $\pm$0.5. We generated datasets containing 400,
200, 100 and 48 points, in which we censored 25, 50 and 75 per cent of the
data. We also performed simulations involving 100 points with a
larger scatter of $\pm$1.
In all cases, the four techniques agreed well (within errors) with
each other and with the original input slope. The Schmitt bisector
generally produced slopes furthest from the OLS bisector slope. 
We conclude
that all four techniques are capable of fitting a single line,
though the survival analysis techniques may have problems in cases of large 
scatter.

\begin{figure*}
\centerline{\psfig{figure=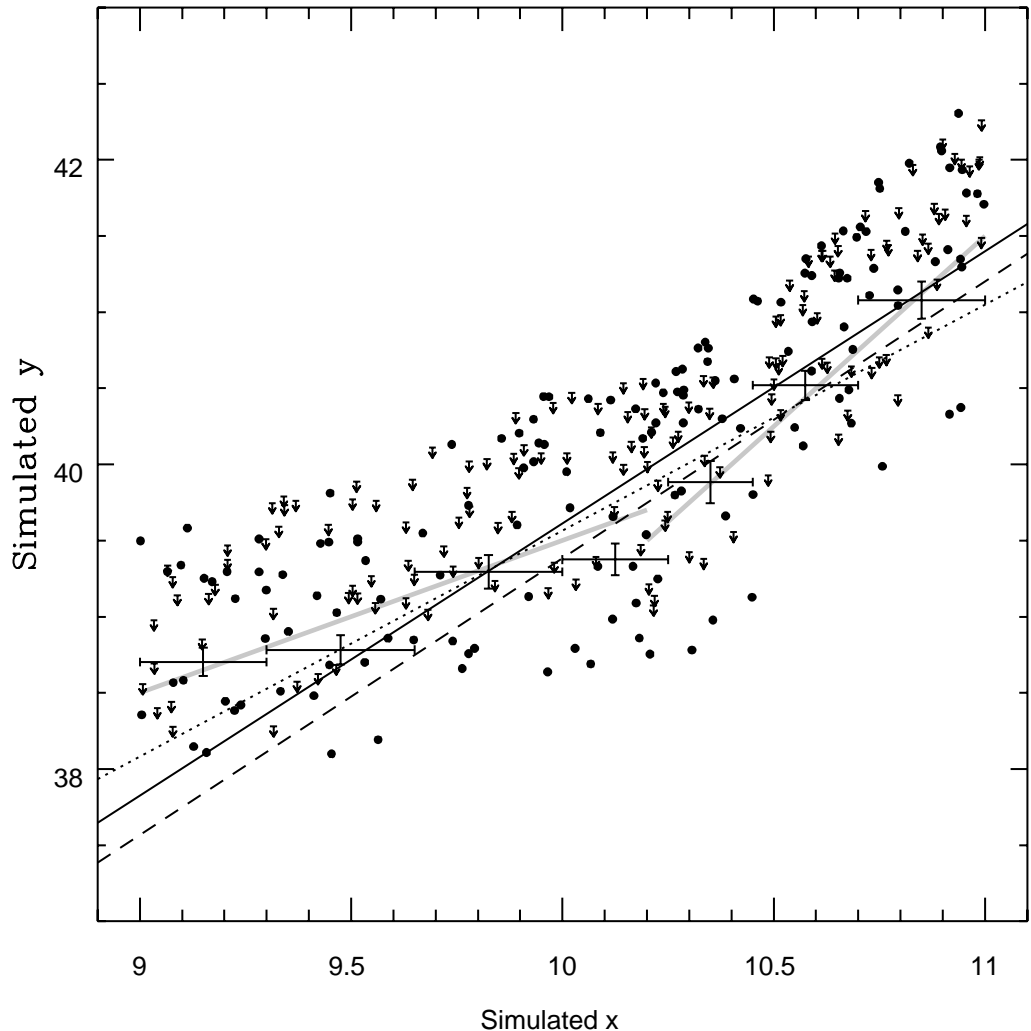,width=6in}}
\vspace{-50mm}
\caption{\label{brPLsim} Simulated censored broken powerlaw data used to
  test alternative fitting methods. The solid line is an OLS bisector fit to
  the underlying, uncensored data, while the dashed line is a Schmitt
  bisector fit to the censored data. The dotted line represents the EM and
  BJ fits. The large crosses show the mean y value for the censored data
  with 1$\sigma$ errors in seven x axis bins, derived using the
  Kaplan--Meier estimator. The two solid grey lines show the original lines 
  used to generate the data points.}
\end{figure*}

However, our data set as a whole does not follow a single linear relationship. 
As can be seen in Figure~\ref{2ndlxlb}, it is probably better 
described as a broken
power law, with a shallow slope below L$_B\simeq$10 and a steeper incline
above. We therefore simulated a new dataset based on
a broken power law similar to that indicated in our data (see Section~\ref{lxlb_total}),
with a gradient of 1.0 over the range x=9.0-10.2, and gradient 2.5 
over x=10.2-11.0. Scatter about each segment was set at $\pm$1 and 
each line was used to generate 150 data points, of which 75 were censored. 
The results of a variety of fits to this simulated dataset are shown
in Figure~\ref{brPLsim}. 

The OLS bisector fit to this dataset gives a slope of $\sim$1.8, as does the
Schmitt bisector which is offset downwards from the OLS bisector line. The
EM and BJ algorithms both have slopes of $\sim$1.49. These results
suggested that the Schmitt bisector does indeed behave, as expected, in a
similar way to the OLS bisector fit, but that both are more influenced by
the steeper line than the shallower. The y/x fits give consistently
shallower slopes, but these may be more representative of
the general spread of points.

To further test the quality of fit, we binned the data from the broken
powerlaw simulation and used the Kaplan--Meier estimator to 
calculate the mean y value in each bin. The binned data are 
shown in Figure~\ref{brPLsim} and follow the original input lines fairly
accurately. All four fit lines deviate from the binned data points at some
point 
on the graph. Both bisector fits deviate quite strongly at low x and are
probably better descriptors of the steeper high x points. The EM and BJ
fits also deviate at low and high x values, but do a rather better job of
describing the overall trend of the binned points across the whole range of
x. 

This result shows clearly that fitting a single line to data which is
better described as the combination of two lines of different slopes will
cause difficulties. The results from the single line
simulations, when considered in conjunction with these results lead us to
further conclude that for our data, which has a high degree of scatter and
is unlikely to be described well by a single line, the Schmitt bisector
should not be used. The EM and BJ algorithms appear likely to give
reasonable estimates of mean trends, but binning the data should provide 
the most accurate picture of the underlying distribution.

\section{Results}
\label{Res2}
Having applied the corrections described in section~\ref{catconv}, we
add a total of 289 early--type galaxies from the three catalogues to our
own. This gives a combined catalogue of 425 galaxies, listed in
Table~\ref{Lxtab2}. When galaxies are listed in more than one catalogue we
choose the final L$_X$ value using the following order of preference: our
results, Beuing \etal (1999), Fabbiano \etal (1992), Roberts
\etal (1991). 
Detections are always preferred to upper limits, regardless of source.  
The
T--type listed in Table~\ref{Lxtab2} is taken from LEDA. The catalogue
contains 24 galaxies which are listed in previous studies as early--type,
but which have LEDA T--types $\geq$--1.5. We exclude these late--type objects
from further consideration.

\subsection{The \lxlbtwo Relation for Early Type Galaxies}
\label{lxlb_total}

\begin{figure*}
\centerline{\psfig{figure=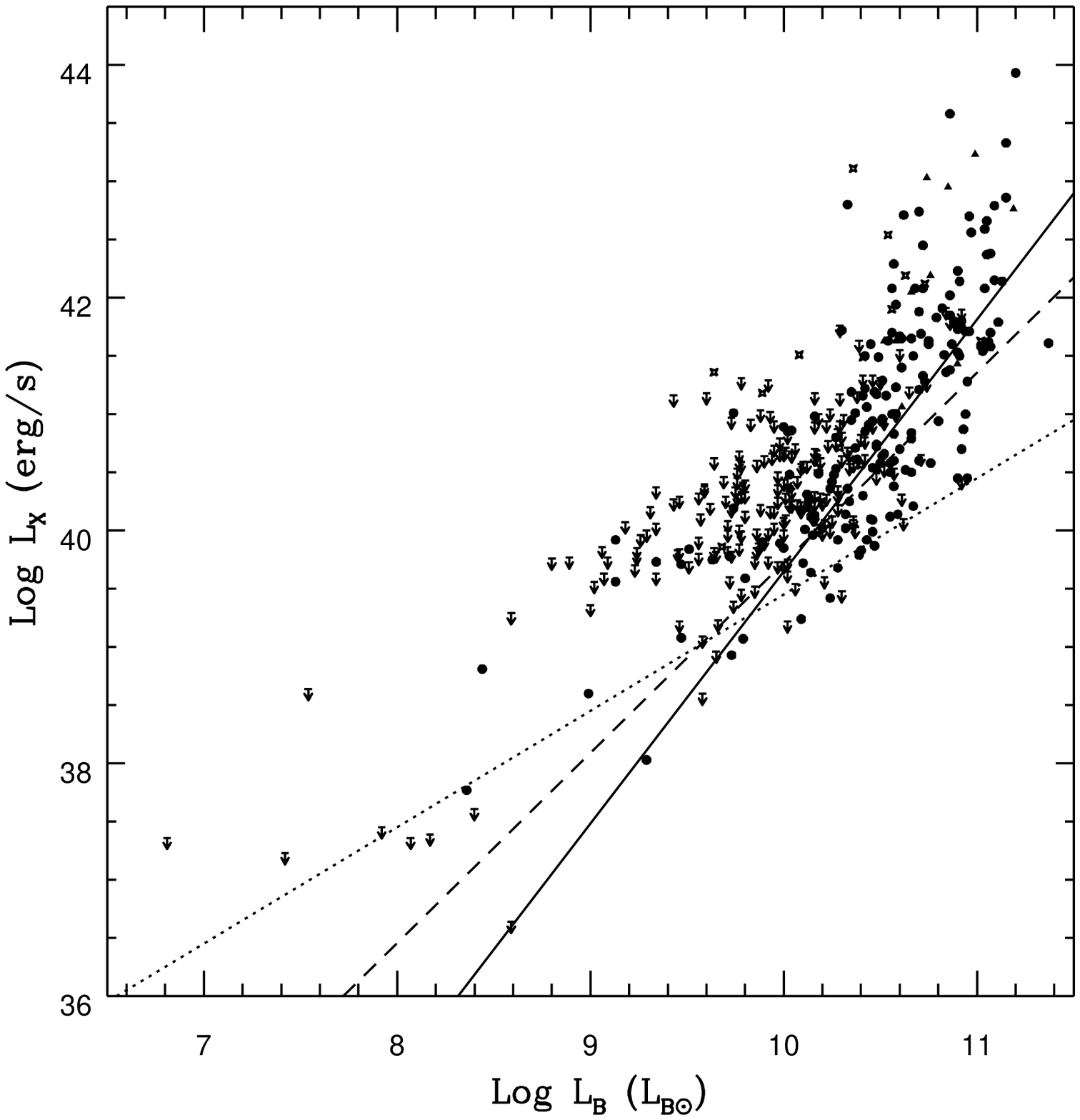,width=6in}}
\vspace{-50mm}
\caption{\label{2ndlxlb} Log L$_X$ vs Log L$_B$ for our full catalogue of
  401 early-type galaxies. Triangles are cluster central galaxies, stars
  AGN and circles all other detections. The lines shown are the best fit
  line to the early--type
  galaxies excluding AGN, BCGs and dwarfs (solid line),
  the best fit to the galaxies excluding all questionable 
  objects (dashed line), and an estimate of the discrete source
  contribution taken from Ciotti \etal (1991).}
\end{figure*}

We have plotted
Log L$_X$ against L$_B$ for the catalogue in Figure~\ref{2ndlxlb}. AGN
(taken from \scite{Veron}) and cluster central
galaxies are likely to have anomalously high X--ray luminosities and are
marked on the plot. Excluding these objects and dwarf galaxies
(L$_B<$ 10$^9$ \LBsol), which are unlikely to be massive enough to retain a
halo of X--ray gas, leaves 359 early--type galaxies of which 184 
have X--ray upper limits. The tests described in Section~\ref{surv} show a
correlation of $>$99.99\% significance. The best fit line from the
expectation and maximization (EM) algorithm is:

\begin{equation}
  Log\thinspace L_X = (2.17 \pm 0.11)\thinspace Log\thinspace L_B + (17.98 \pm 1.12)
\end{equation}

and from the Buckley--James (BJ) algorithm:

\begin{equation}
  Log\thinspace L_X = (2.28 \pm 0.12)\thinspace Log\thinspace L_B + 16.80
\end{equation}

The standard deviations about the two regression lines are
$\sigma_{\sc em}$=0.69 and $\sigma_{\sc bj}$=0.68 respectively.

These values are in fairly good agreement with a number of previous
estimates (\pcite{Beuingetal99}; \pcite{donnellyfaberoconnell90};
\pcite{whitesarazin91}), but differ from those of
\scite{brownbreg98} who found a slope of $\sim$2.7 using a small sample
of optically bright galaxies. 

X--ray emission from discrete sources is expected to produce a lower bound
to the distribution of galaxies in Figure~\ref{2ndlxlb}. We
discuss the question of discrete source emission in detail in
Section~\ref{disc_sec}, but at this stage we note that our points are
reasonably consistent with the estimate of discrete source emission made by
\scite{ciotti91}.

Previous work with group galaxies (\pcite{Steve}) has shown that the
properties of central dominant group galaxies are substantially different
from those of normal group members and field galaxies. The X-ray luminosity 
of these dominant galaxies is in fact more closely correlated with the
properties of the group as a whole than with the optical luminosity of the
galaxy (\pcite{HelsdonPonman01}). Temperature profiles of X-ray bright groups
suggest that these objects are at the centres of group cooling flows, which 
explains their overluminosity compared to 
non-central galaxies. The Brown \& Bregman sample contains a large number
of group--dominant galaxies (\pcite{Steve}), which may account for the high
slope of their best fit \lxlbtwo relation.

Group--dominant galaxies can be identified by their position at the centre
of the group X--ray halo. Unfortunately, since part of our catalogue is drawn
from the 
literature, we are unable to carry out identifications in this way. However,
group--dominant galaxies are usually the most massive and luminous object in 
the group. In order to remove any bias produced by these dominant galaxies,
we excluded all brightest group galaxies (BGGs) and then fitted the
remaining data. The majority of BGGs are selected using the catalogue of
groups by \scite{Garcia93}. However, 48 of our galaxies lie beyond the
Garcia redshift limit (5,500 km s$^{-1}$), and in these cases we are forced
to identify BGGs using other catalogues. Using NED, we were able to check
each galaxy for membership of the catalogues of \scite{Abelletal89},
\scite{Whiteetal99}, \scite{Hickson82} and \scite{M98j}. As White
\etal do not list the BGG of each group, we have identified them based on
the apparent magnitudes given in NED. 

In order to check for other objects which might bias the fits, we also 
used NED to check all galaxies with log \lxlb $>$ 31.5 for unusual
properties. A surprising number of these 
objects show potential problems. For example, we found several probable AGN 
not identified in \scite{Veron} (\eg NGC 3998, NGC 4203, NGC 7465). 
Excluding all BCGs,
BGGs, AGN and dwarf galaxies, leaves a total of 270 objects. Fitting \lxlbtwo
for this reduced sample lowers the slope of the best fit lines
significantly, to 1.98$\pm$0.13 (EM) or 2.17$\pm$0.15 (BJ), with
$\sigma_{\sc em}$=0.69 and $\sigma_{\sc bj}$=0.70. This
change demonstrates the influence of BGGs on the L$_X$:L$_B$ relation. 

As a final precaution we also fit a very conservative subsample, from which 
we have removed not only all AGN, BCGs, BGGs and dwarf galaxies, but also
all objects which lie at a distance $>$70 Mpc, to avoid including
misclassified galaxies. We also remove the anomalous galaxies NGC 5102 and
NGC 4782 from this conservative subsample. NGC 4782 has an unusually high
L$_B$, and the B magnitude given 
for it in Prugniel \& Simien (1996) disagrees with those in LEDA and NED by
$>$1 magnitude. NGC 5102 is a relatively small E--S0
galaxy (Log L$_B$ = 9.29 L$_{\odot}$) with an exceptionally low X--ray
luminosity (Log L$_X$ = 38.03 erg s$^{-1}$). It
is thought to have undergone an episode of star formation a few
10$^8$ years ago (\pcite{BicaAlloin87}). Although during and shortly after
the starburst we might expect to observe an enhanced L$_X$ compared to
L$_B$ (\pcite{Readponman98}), galactic wind models predict that the
starburst can remove all gas from the galaxy, leaving it significantly
underluminous until the halo is rebuilt (\pcite{ciotti91};
\pcite{pellegriniciotti98}). The B--band luminosity will also be enhanced
by the population of young stars produced by the starburst, making the
position of such an object on an L$_X$:L$_B$ plot even more aberrant. 

Removing all of these objects 
reduces our dataset to 246 galaxies, of which 159 have X--ray upper
limits. This lowers the slope of the best fit lines considerably, to
1.63$\pm$0.14 (EM) and 1.94$\pm$0.17 (BJ), with $\sigma_{\sc em}$=0.60
and $\sigma_{\sc bj}$=0.62. The difference in results
between the two fitting methods in this case is large, particularly as the
1$\sigma$ error regions do not overlap. As mentioned in
Section~\ref{surv}, the only difference between the two techniques is the
assumed underlying distribution of points. As the EM method assumes the distribution
to be normal, we tested the distribution of detections (87 points) for
normality, using the algorithm AS 248 (\pcite{DavisStephens89};
\pcite{Stephens74}) which provides several measures of goodness of
fit. These tests showed that the detected points were normally distributed
about the best fit line at 50-60\% significance. This is not a strong
confirmation of the normality of the data, but is also not poor enough to
rule out a normal distribution. 

It is notable that the agreement between the two fitting algorithms
worsens as the fraction of upper limits in the data increases. Our complete 
catalogue has $\sim$50\%, and the fits are in reasonable agreement, whereas 
our conservative subsample has $\sim$65\% upper limits and shows poor
agreement. \scite{Isobeetal86} simulate fits to datasets containing 30
points, of which $\frac{2}{3}$ are upper limits, and produce acceptable
results, but their data does not appear to have as large a degree of
scatter as ours. It seems likely that our conservative subsample is rather
poorly constrained, and is perhaps not well modeled by a normal
distribution. This suggests that the BJ method is the more reliable 
in this case.

\begin{figure*}
\centerline{\psfig{figure=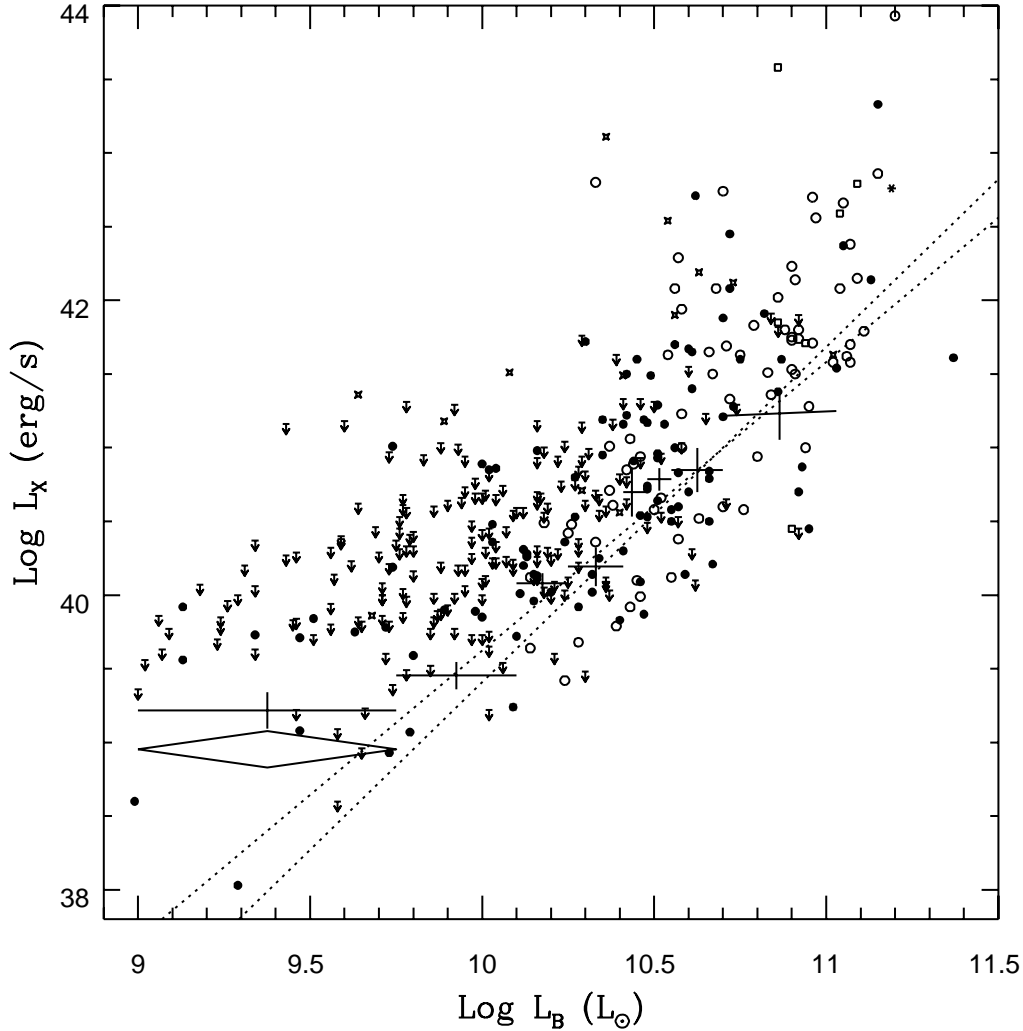,width=6in}}
\vspace{-50mm}
\caption{\label{binnedlxlb} Our catalogued data L$_X$ and L$_B$ data with
  mean L$_X$ values for eight bins. Open circles are BGGs, triangles are
  BCGs, stars are AGN, filled circles are detected normal galaxies and
  arrows are upper limits. The large crosses show the mean L$_X$ in eight
  bins, calculated using the Kaplan--Meier estimator and excluding AGN,
  BCGs, BGGs, dwarfs, galaxies at distances $>$70 Mpc, NGC 5102 and NGC
  4782. The diamond shows the mean L$_X$ in the lowest bin, corrected to remove
  the expected contribution from discrete sources (see
  Section~\ref{subtracted}). The dotted lines are EM (shallower) and BJ 
  (steeper) fits to the same data.}
\end{figure*}

Even excluding BGGs, there is some evidence of a change in the slope of the 
\lxlbtwo relation above L$_B\simeq$10 L$_{B\odot}$.
To see how this apparent change in slope affects our fits, we binned the
very conservative sample and
calculated the mean L$_X$ in each bin. These are plotted in
Figure~\ref{binnedlxlb}. The bins clearly follow a general trend, but
at low L$_B$, the gradient becomes shallower. We also defined a new sample
of data which excludes AGN, BGGs, BCGs, galaxies with distances $>$70 Mpc
L$_B<$10, NGC 5102 and NGC 4782. This sample should have had most points which
are likely to bias the fit removed, and with L$_B>$10 it should model the
steeper section of the relation. EM and BJ fits to this sample give slopes
of 1.96$\pm$0.25 and 2.28$\pm$0.29 respectively, with standard deviations
about the fits in both cases of $\sigma$=0.58.
Both fits seem to do a
reasonably good job of matching the binned data points at L$_B>$10, with
the EM fit being slightly closer to the points at high and low L$_B$.

\subsection{Potential sources of bias}
\label{Res1}

Our catalogue is made up of X--ray luminosities which can be split in to
three main categories; those which we have calculated based on pointed {\it
  ROSAT} PSPC data, those which are based on {\it ROSAT All--Sky Survey}
data, and those which are based on pointed {\it Einstein} IPC data. Clearly 
it is important to examine possible biases which may arise from this
combination of data.

\scite{Sansom00} have carried out a {\it ROSAT} study of 52 galaxies with
optical fine structure. In order to check the accuracy of their own
analysis of PSPC pointed data, they compare their own count rates with those of
\scite{Beuingetal99}. For the majority of their sample both analyses are in 
agreement, but they note that in three cases the count rates differ by more 
than a factor of two. The objects concerned are NGC 7626, in the Pegasus I
cluster, NGC 3226 which has an X--ray bright neighbour, and NGC 4203 which
is close to an unrelated X--ray source. The inclusion of the neighbouring
sources in the Beuing \etal analysis for the latter two cases is caused by
the short exposure times (typically $\sim$400 s) of RASS
observations. Although extraction radii for detected galaxies were based
on surface brightness profiles, low numbers of counts may cause close pairs
of sources to be blurred together, appearing as a single object.

A similar but perhaps more serious problem occurs in cases where the target 
galaxy is surrounded by X--ray emission from a group or cluster halo. In
these cases, Beuing \etal calculate luminosities for those galaxies which
clearly stand out from the emission or are at the center of emission which
is reasonably symmetric around them. Galaxies which stand out from the
background emission may have overestimated luminosities, owing to the
inclusion of emission from that part of the group/cluster halo lying along
the line of sight. However, this is true of most luminosity estimates for
galaxies in such environments, and as the galaxy clearly stands out against 
its surroundings, it 
seems fair to assume that its own emission dominates. On the other hand,
it seems likely that galaxies in the centres of groups and clusters will have 
seriously over-estimated luminosities, due to the inclusion of the majority 
of the surrounding halo emission. Beuing \etal exclude cluster dominant
galaxies from their fitting, but not group--dominant galaxies, which may
steepen the slope of their \lxlbtwo relation. 

Despite the corrections described in Section~\ref{catconv}, we are almost
certainly including some data from Beuing \etal which are biased by
inclusion of extraneous sources or 
group emission. However, we perform fits which exclude BGGs, and may
therefore expect to remove the majority of the most biased points. It is also
worth noting that Beuing \etal calculate upper limits on X--ray luminosity
using a fixed aperture 6 optical half--light radii in diameter, and do not
use upper limits for galaxies embedded in bright cluster
emission. 

Luminosities calculated from {\it Einstein} data are generally based on
considerably larger numbers of counts than those taken from the RASS. They
should therefore be somewhat less likely to suffer from the problems
described above. Unfortunately the poorer spatial resolution of the IPC
compared to the PSPC makes confusion of close sources more likely,
particularly if the sources are relatively faint. Our comparisons in
Section~\ref{catconv} show that there is no major systematic offset, but
there are still likely to be individual galaxies which have been
over--estimated.

In our own analysis of PSPC pointed data we have attempted to avoid these
problems where possible. Confusion between close sources should be minimal, 
as we work with considerably larger numbers of counts. We have attempted to 
remove at least a part of any contaminating group or cluster emission where 
possible, reducing the degree to which group and cluster gas biases the
\lxlbtwo relation. However, without two dimensional fitting of the surface
brightness profile of the group and galaxy it is not possible to completely 
remove this contamination, so we must expect to over--estimate some of the
luminosities.

\begin{figure*}
\centerline{\psfig{figure=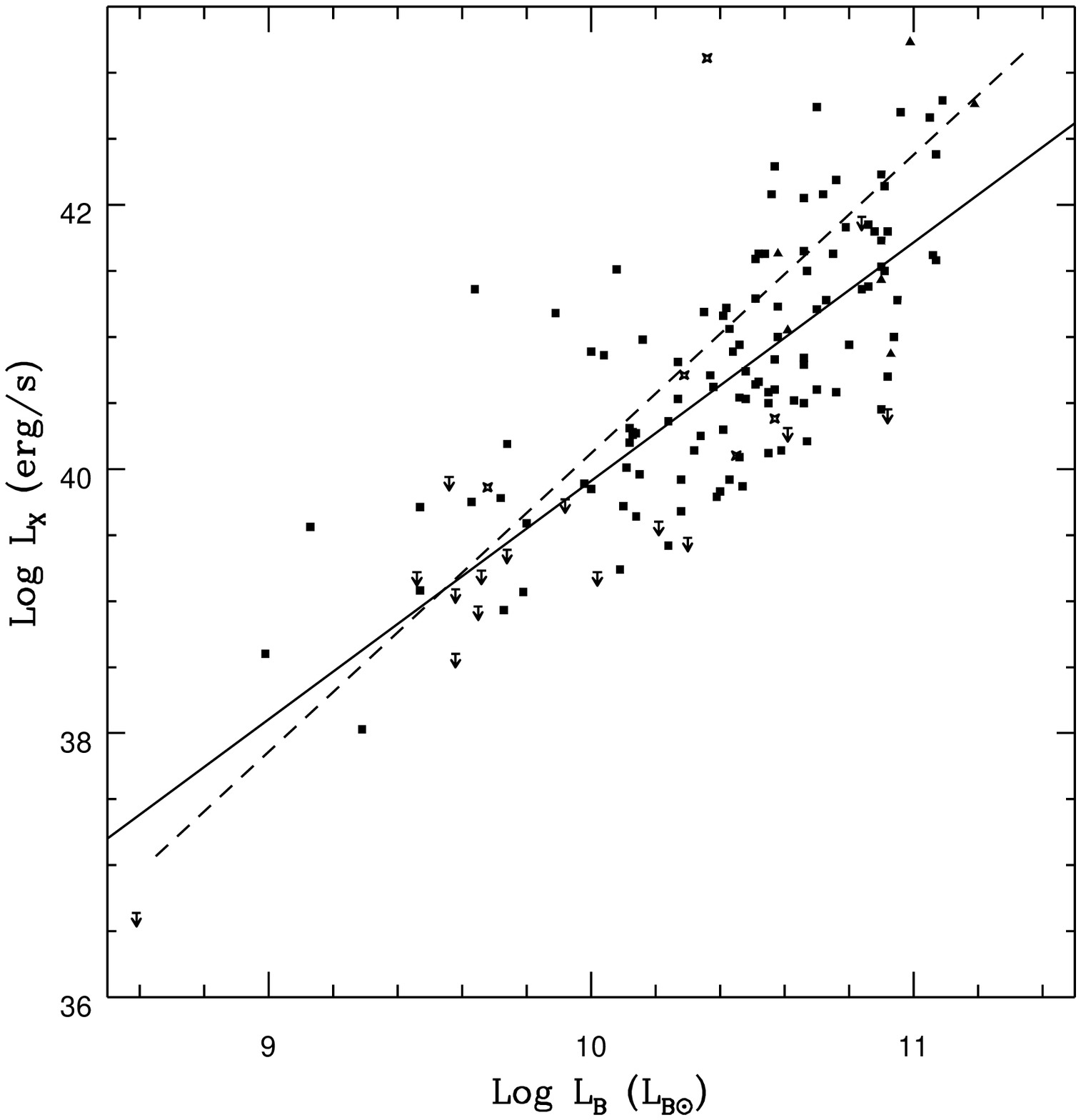,width=6in}}
\vspace{-50mm}
\caption{\label{1stlxlb} Log L$_X$ vs Log L$_B$ for our sample of 136
  early--type galaxies from {\it ROSAT PSPC} pointed
observations. 
Filled squares
  are early--type galaxies, triangles represent
  cluster central galaxies and stars AGN. The solid line is our best fit to 
  the data, excluding AGN, dwarfs and cluster central galaxies. 
The dashed line is
  the Beuing \etal best fit relation.}
\end{figure*}

In order to examine our data for possible biasing effects
we have fitted an L$_X$:L$_B$ relation for a subset of the catalogue,
made up of those galaxies whose X--ray luminosities are the product of our
own analysis of {\it PSPC} data. Figure~\ref{1stlxlb} shows a plot of Log L$_X$
vs Log L$_B$ for this sample.  
For the complete subset, the statistical tests described
in section~\ref{surv} show a probability $>$99.99\% that a correlation
exists, and give slopes of 1.73$\pm$0.12 (EM) and 1.71$\pm$0.13 (BJ)
respectively. The standard deviations about the two regressions are
$\sigma_{\sc em}$=0.74 and $\sigma_{\sc bj}$=0.69. 

As discussed in Section~\ref{lxlb_total}, fitting a line to the complete
sample does not provide a good estimate of 
the true L$_X$:L$_B$ relation for the sample, as there are a number of
unusual objects included. Removing cluster central galaxies, AGN
and dwarf galaxies steepens the slope to 1.81$\pm$0.15 (EM) or
1.79$\pm$0.15 (BJ), with $\sigma_{\sc em}$=0.61 and $\sigma_{\sc
  bj}$=0.59. The EM fit is shown as the solid line in 
Figure~\ref{1stlxlb}.

For comparison the L$_X$:L$_B$ relation of
\scite{Beuingetal99}, which has a slope of 2.23$\pm$0.13, is shown.
It is clear that the Beuing \etal line is not a particularly 
good fit to the data. However, our fitted slopes are similar to the slope
found by \scite{fabbianokimtrinchieri92} for their sample of elliptical
galaxies observed with {\it Einstein}. We 
believe that these differences in slope are caused by the different
analysis strategies adopted for the three samples, and that the steeper
slope of the Beuing \etal data may be caused by cases of over--estimation
of L$_X$, as discussed above.

\subsection{The Discrete Source Contribution}
\label{disc_sec}

The X-ray emission from early-type galaxies is thought to be produced by a
combination of sources. These can be generalized into two categories; hot
gas and discrete sources. Discrete sources (\eg X-ray binaries, individual
stars, globular clusters) are essentially stellar in origin and so the
total X-ray luminosity from these sources should scale with L$_B$. This can 
seen in the \lxlbtwo relation of Beuing \etal (1999), 
which at low L$_B$ agrees well with discrete 
source estimates with slope unity. However, the normalization of these
discrete source estimates is not well defined -- those quoted in Beuing
vary over at least an 
order of magnitude, and only the highest is ruled out by that data set.

Most previous estimates of the discrete source contribution to L$_X$
(hereafter \Ldscr) are based on a small number of relatively nearby
objects. For example, \scite{trinchierifabbiano85} base their estimates on
{\it Einstein} observations of M31, \scite{fjt85} use Centaurus A, while
\scite{irwinsarazin98} use M31 and NGC 1291. Estimates based on early-type
galaxies are rare, as it is difficult to separate a discrete source
component from the overall emission. One simple method to avoid this
problem is employed by \scite{ciotti91}, who fit a slope unity line to the
lower envelope of data from \scite{CFT87}. This gives an estimate of
log(\Ldscr/L$_B$) = 29.45 erg s$^{-1}$ (using our pseudo-bolometric bandpass
and model). This value has 
been shown to be a good estimate of the lower bound of the Beuing \etal
sample, and is also a reasonably good match to our data. However, this does not
necessarily imply that the value is a good estimate of the mean \Ldscr. To
produce more accurate estimates we need either a large sample of data from
late-type galaxies which have little or no hot gas emission, or much more
detailed spectral studies of early-type objects.

\subsubsection{X-ray emission from late-type galaxies}
Late-type galaxies are known to be sources of X-ray emission, though
not of the same magnitude as elliptical and S0 galaxies (\eg
\pcite{fabbianokimtrinchieri92}). Early studies
(\pcite{fabbianotrinchieri85}) showed that there is a strong correlation
between the X-ray and optical emission, giving rise to an \lxlbtwo
relation similar to that observed for early-type galaxies. However, in
late-type galaxies this relation has a much shallower slope than in
early-types. Most studies find this slope to be \gtsim 1.

The most common explanation for this relation is that the X-ray emission
observed is produced mainly by X-ray binaries and hot stars. As these
sources are stellar in origin, their numbers should be directly related to
the optical luminosity of the galaxy, and the \lxlbtwo relation for spirals
should have a slope of $\simeq$ 1. Emission from other sources, such as
hot gas, may not be so directly linked to stellar populations. If spiral
galaxies contain significant amounts of hot gas as well as discrete
sources, we would expect to see an \lxlbtwo relation for with a slope $\geq$ 1.
  
Detailed spectral studies of the X-ray emission from
nearby spiral galaxies (\eg \pcite{turneretal97}; \pcite{RPS97};
\pcite{Ehleetal98}) have shown that such a hot gas component is present in
some cases.
Using a large sample of galaxies observed with {\it Einstein},
\scite{KimFabbTrin92} showed that this ISM component was mainly associated
with early--type (Sa) spirals, and that there was a succession of spectral
properties with morphology. Elliptical and E/S0 galaxies were mainly
dominated by gaseous emission, S0 galaxies had a somewhat harder spectrum,
Sa spirals were harder still with the hard component dominating, and
late--type spirals showed little sign of a hot ISM. This points toward the
hot gas being associated with the bulge of the galaxy; Sb and Sc galaxies
have small bulges and little or no hot gas, whereas ellipticals are
essentially all bulge, and have large gaseous halos. 

More recent studies have confirmed the lack of significant halos around
spiral galaxies. \scite{bensonetal99} used {\it ROSAT} observations
of three massive edge-on spiral galaxies to look for large scale
extended emission predicted by galaxy formation models to arise as hot gas
cools to form the galaxies' disks. They
found no evidence for X--ray halos of the extent seen around early--type
galaxies. Studies of diffuse emission within or near spiral galaxies
suggest that hot gas does not extend far beyond the stellar body of the
galaxy except in the case of starburst galaxies (\pcite{RPS97}). We have
avoided all such galaxies, as the contribution to the X--ray emission from
active star--formation and the associated galactic winds would seriously affect
our results.  

To define an L$_{dscr}$:L$_B$ relation for spirals we tried two approaches. The
first was to search the literature for attempts to separate gaseous and
discrete emission in spiral galaxies. The second was to obtain a large
sample of spiral galaxies observed in X-rays and split this into subsamples 
by morphological type. The hierarchical merger scenario implies that galaxies
with small bulges should have minimal X-ray emission from hot gas, so there 
should be a trend for a lower and less steep \lxlbtwo relation for
later-type spirals. 

\subsubsection{Nearby spiral galaxies sample}
We have collected L$_X$ values for 13 spirals, and 
normalised them to bolometric
luminosities. The following list gives details of
these sources: 

\begin{itemize}
\item Nine galaxies from \scite{RPS97}. The objects chosen are those which are 
  not listed as starburst galaxies in the paper, NGC 55, NGC 247, NGC 300,
  NGC 598, NGC 1291, NGC 3628, NGC 3628, NGC 4258 and NGC 5055. The X-ray luminosities
  given are for emission from the galaxies after any resolved point sources 
  had been removed.
\item M83, from \scite{Ehleetal98}. The $L_X$ value given is for the harder 
  of two diffuse emission components fitted, thought to represent
  unresolved discrete sources in the disk and bulge. Resolved point sources 
  were removed before fitting.
\item Centaurus A (NGC 5128), from \scite{turneretal97}. The $L_X$ value given is for
  the 5 keV component of a two temperature Raymond \& Smith plasma model
  fitted to the diffuse emission from the galaxy. Regions contaminated by
  the nucleus and associated jet were removed, but some point source
  emission was included.
\item NGC 4631, from \scite{fabbianotrinchieri87}. The value used is that
  given for a soft (0.2-0.8 keV) component associated with the disk of this 
  galaxy.
\item The bulge of M31, from \scite{irwinsarazin98a}. 
\end{itemize}

\begin{figure*}
\centerline{\psfig{figure=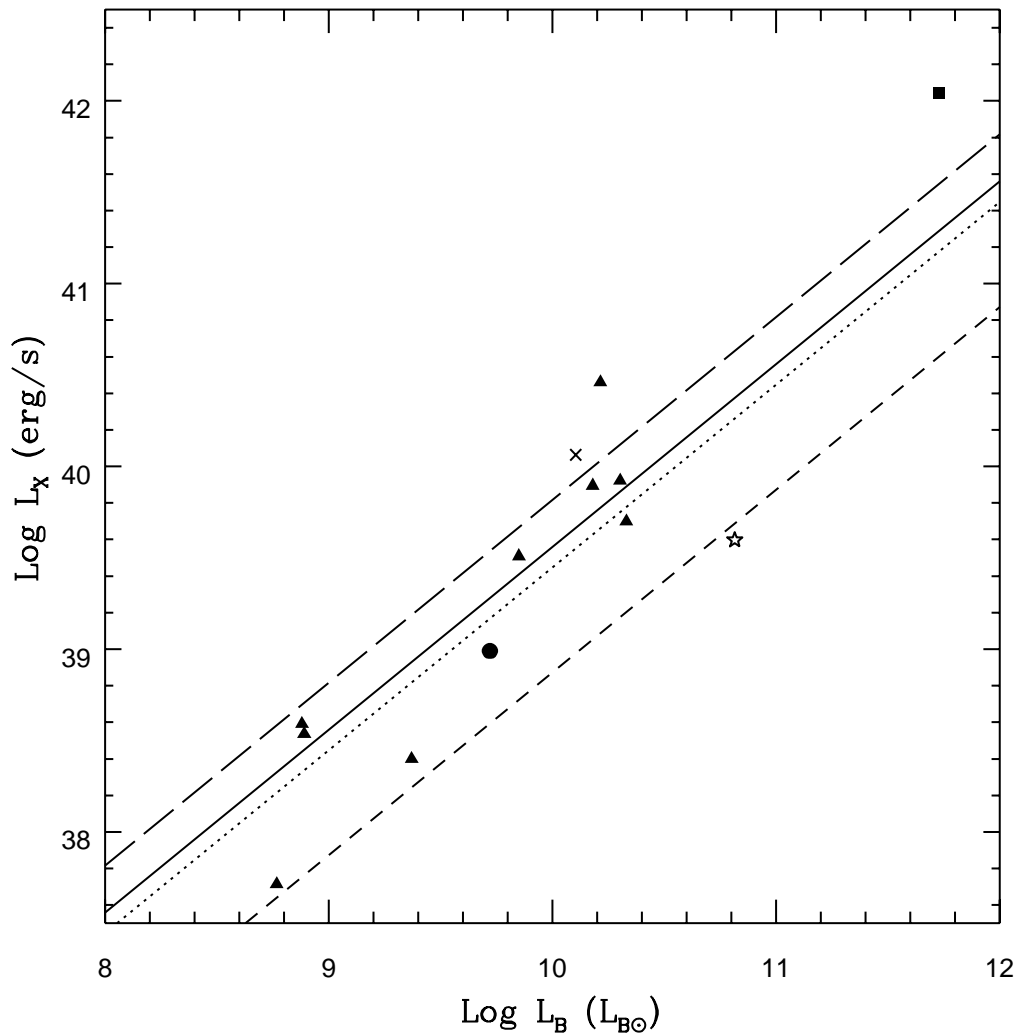,width=6in}}
\vspace{-50mm}
\caption{Plot of $L_X$ vs $L_B$ for a sample of nearby late-type galaxies.
  Solid
  triangles represent data from Read, Ponman \& Strickland (1997), the
  square M83, the cross
  NGC 4631, the star Centaurus A, and the circle the bulge of M31. The solid 
  line is the best fit slope unity line, with the other lines showing
  estimates of $L_{dscr}$ from the literature. The dotted line is the
  estimate of Ciotti \etal (1991), the short dashed is taken from Forman \etal
  (1985) and the long dashed is from Canizares \etal (1987)}
\label{nearbyspirals}
\end{figure*}

In most of the cases listed above, we have selected the component of
emission which is most likely to represent the discrete sources in each
galaxy, and excluded components corresponding to gaseous emission. However, 
we have also excluded a number of resolved point sources, which could
be a part of the discrete source population. To be resolved by the
instruments used in these studies, the point sources must be highly
luminous. At worst, this suggests that they might be AGN or bright
transient sources. At best they could be unusually powerful LMXBs, or
possibly black hole binaries. We have decided to exclude these sources to
avoid the possibility of contaminating the sample with emission from
objects which are not part of the population in which we are interested. 

The results are shown in Figure~\ref{nearbyspirals}. {\sc iraf} survival analysis
tasks were then used to fit lines to these points, both with a fixed
slope of unity and with the slope allowed to vary. Using the Kaplan-Meier
estimator, we found the intercept of the slope unity line to be 29.56 $\pm$
0.13. Fitting of a variable slope line with the EM algorithm gave a slope
of 1.21 $\pm$ 0.15 and an intercept of 27.51 $\pm$ 1.44. The fixed slope
line is plotted on Figure~\ref{nearbyspirals}, as well as three estimates
of discrete source emission taken from \scite{ciotti91}, \scite{fjt85} and
\scite{CFT87}. Our line agrees within errors with that of Ciotti \etal.

\subsubsection{Morphologically defined samples}
\label{morphsec}

\begin{table}
\hspace{-5mm}
\begin{center}
\begin{tabular}{|l|c|c|c|c|}
\hline
Group & N & Unit slope & \multicolumn{2}{c|} {Variable slope}\\
\cline{3-5}
 & & intercept & slope &\ intercept\\
\hline
Sa & 90  & 30.59 $\pm$ 0.11 & 2.14 $\pm$ 0.31 &\ 19.07 $\pm$ 3.07\\
Sb & 74  & 30.22 $\pm$ 0.09 & 1.14 $\pm$ 0.29 &\ 28.77 $\pm$ 2.90\\
Sc & 98  & 30.12 $\pm$ 0.06 & 1.38 $\pm$ 0.16 &\ 16.35 $\pm$ 1.60\\
\hline 
\end{tabular}
\end{center}
\caption{Slopes and intercepts of four morphological subsamples selected
  from Fabbiano, Kim \& Trinchieri (1992)}
\label{sp_tab1}
\end{table} 

\begin{figure*}
\centerline{\psfig{figure=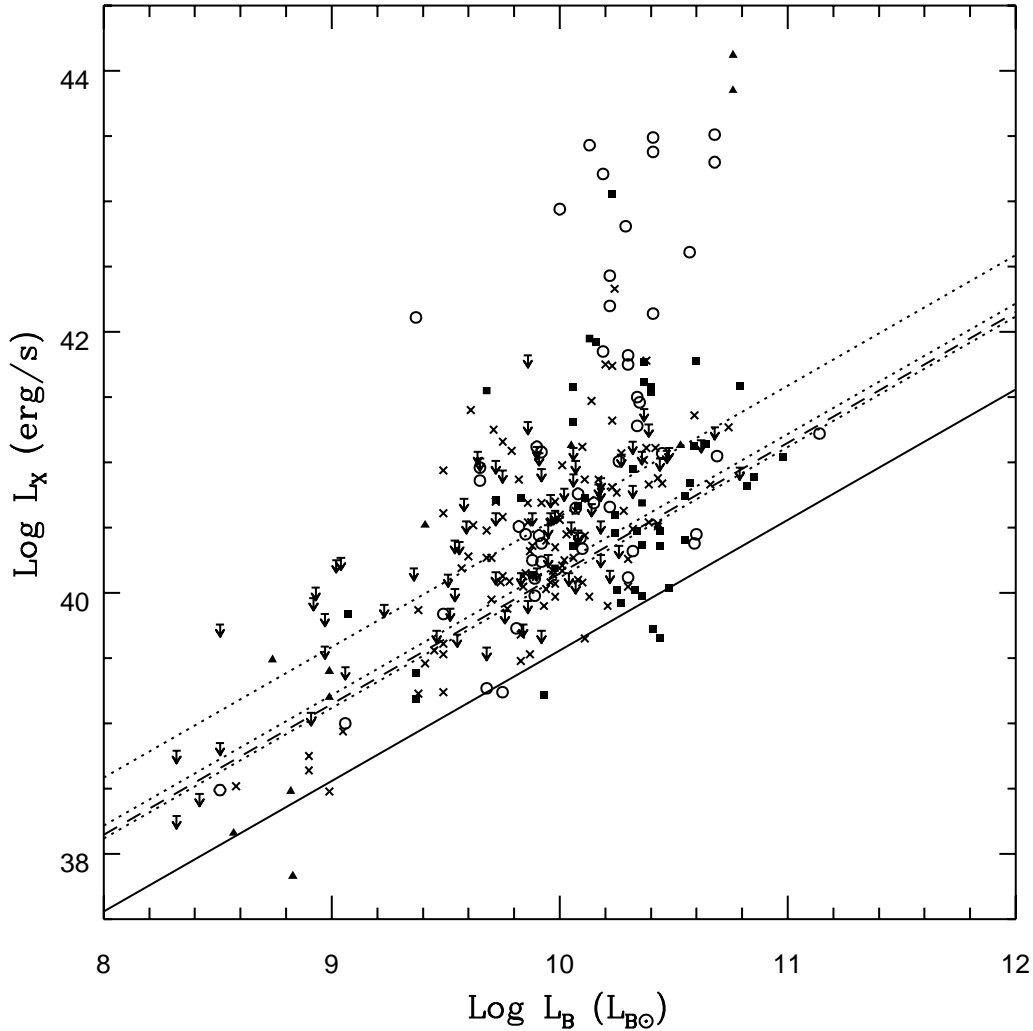,width=6in}}
\vspace{-50mm}
\caption{Plot of $L_X$ vs $L_B$ for morphological subsets of late-type
  galaxies from Fabbiano, Kim \& Trinchieri (1992). Circles are Sa, squares 
  Sb, crosses Sc, and triangles Sd and Irregular galaxies. Upper limits
  from all classes are shown as arrows. The solid line is the same as that
  shown in Figure~\ref{nearbyspirals}, while the dotted lines are slope unity fits to the
  four subclasses. The dashed line corresponds to the slope unity fit to
  the data below Log L$_B$ = 9.9 \LBsol. See text for further details.}
\label{Morphspirals}
\end{figure*}

Working with the large spiral sample of Fabbiano \etal (1992) we define
three morphological subsets; Sa, Sb and Sc. Results from fits to the
\lxlbtwo properties of these subsets with fixed (unity) and variable slopes 
are shown in Figure~\ref{Morphspirals} and listed in Table~\ref{sp_tab1}.
It can be seen that there is a distinct difference between the earlier-type
spirals in group 1, and the later types in groups 2 and 3.
Therefore, it seems that these results support the idea that X-ray gas
luminosity is correlated with bulge size, though the effect is not large.

As a check of this result we have also carried out fits to samples of
spiral galaxies 
taken from the catalogue of \scite{Bursteinetal97}. This catalogue,
although it contains a larger number of galaxies than that of Fabbiano
\etal, is dominated by upper limits and uses an average of three spectral
models to convert count rates to fluxes. We have not therefore converted it 
to our waveband and model, but have instead simply compared general trends
in the results with those we find using the Fabbiano \etal data. Fitting
lines of unit slope to Sa, Sb and Sc subsamples we find a similar trend in
relative normalisation; the Sa sample has an intercept significantly above
either of the other subsets.

The line fits to the Fabbiano \etal data and the data values themselves can
be seen in Figure~\ref{Morphspirals}. For comparison the fit to the nearby
spiral data discussed above is also 
shown. It is clear that even the lowest of the fits to the
\scite{fabbianokimtrinchieri92} data is considerably higher than that to the
nearby galaxies, presumably owing to the removal of point sources and (in
some cases) gaseous emission from the nearby spiral data.

In Figure~\ref{Morphspirals}, it can be seen that as $L_B$ increases, the data points for
all the subclasses diverge more and more from the slope unity lines. We
therefore decided to split the sample into two new subsets. These are the
low luminosity (Log $L_B \leq$ 9.9 \LBsol) and high luminosity (Log $L_B >$
9.9 \LBsol)
subsets. Line fits for each are shown in Table~\ref{sp_tab2}.

\begin{table}
\begin{center}
\begin{tabular}{|l|c|c|c|c|}
\hline
Group & N & Unit slope & \multicolumn{2}{c|} {Variable slope}\\
\cline{3-5}
 & & intercept & slope &\ intercept\\
\hline
Low $L_B$ & 115 & 30.15 $\pm$ 0.06 & 1.01 $\pm$ 0.13 &\ 30.34 $\pm$
1.18\\
High $L_B$ & 164 & & 2.03 $\pm$ 0.38 &\ 20.13
$\pm$ 3.93\\
\hline 
\end{tabular}
\end{center}
\caption{Slopes and intercepts for the high (Log $L_B >$ 9.9 \LBsol) and low luminosity subsets.}
\label{sp_tab2}
\end{table}

These figures show clearly that there is a large difference in the
\lxlbtwo relation for low and high luminosity spirals. The high $L_B$
subset has a slope similar to that found for elliptical galaxies, while
the low $L_B$ sample slope is very close to 1. The slope unity
fit to the low L$_B$ sample is  shown in Figure~\ref{Morphspirals} as a dashed line.

\subsubsection{\Ldscr from early-type galaxies}

In order to directly measure \Ldscr from early-type galaxies, it is
necessary to distinguish between emission from hot gas and the contribution 
of the discrete source population. Whereas X-ray bright galaxies are
usually fit using a single component MEKAL or Raymond \& Smith model with
kT$\sim$1 keV, X-ray faint galaxies have been shown to be better fit by
two component models (\pcite{fabbianokimtrinchieri92};
\pcite{Pellegrini94}). These consist of a high temperature (kT
$\sim$ 10
keV) component generally associated with X-ray binaries and a low
temperature component with kT$\sim$0.2 keV. A number of possible sources
for this low temperature component have been suggested
(\eg \pcite{irwinsarazin98}), but LMXBs again seem to be the most likely
source (\pcite{irwinsarazin98a}; \pcite{IrwinSarBreg00}). The advent of 
{\it Chandra} has made it possible to resolve significant numbers of point
sources in nearby galaxies. Observations of NGC 4697
(\pcite{SarazinIrBreg00}) and NGC 1553 (\pcite{Blantonetal01}) reveal
considerable numbers of point sources with hard spectra. Blanton \etal show
that, at least in the case of NGC 1553, the emission 
from resolved point sources is best fit using a model which includes a low
temperature component. From the Blanton \etal results we estimate that the
total flux from discrete sources, excluding the AGN, is
8.58$\times$10$^{-13}$ erg s$^{-1}$ cm$^{-2}$ in the 0.3-10 keV band. Using
our distance and L$_B$ for NGC 1553 gives \Ldscr = 29.44. As we do not have 
the exact details of the Blanton \etal best fit model, we cannot convert
this to our own model and waveband, but any correction should be small, as
the {\it Chandra} waveband extends to much higher energies than that of
{\it Einstein} or {\it ROSAT}. Assuming a 20\% conversion factor produces
\Ldscr = 29.52, very similar to our other estimates. However, both NGC 1553
and NGC 4697 have low X--ray luminosities, and relying purely on the lowest
luminosity galaxies for measurements of \Ldscr may 
be unwise. \scite{irwinsarazin98} note that small fluctuations in the
discrete source populations in these objects could cause a large degree of
scatter in L$_X$, as their total X-ray luminosities are small. This is
confirmed by comparison of the luminosity functions of the point source
populations of four galaxies observed by {\it Chandra}
(\pcite{Kregenowetal01}). Theses show differences in \Ldscr of a factor
$\geq$ 4 between the galaxies. Clearly estimates based on large samples are 
likely to be more reliable.

With high quality {\it ASCA} data, it is possible to fit high luminosity
galaxies using both a 1 keV
gaseous halo and a high temperature discrete component (\eg
\pcite{Matsumotoetal97}). The most recent study of this sort
(\pcite{Matsushita00}) fits a 10 keV bremsstrahlung model to 27 galaxies,
excluding those which show signs of harboring low luminosity AGN,
producing a value of Log \Ldscr = 29.41 L$_{B\odot}$ 
(converted to our passband and
model). Given the quality of the data used, this is probably the most
reliable value available from studies of
elliptical galaxies. Its only drawback is that it does not take into
account the effects of a low 
temperature component in LMXB emission. Such a component is unlikely to be
detected by {\it ASCA}, which has a relatively low collecting area and poor 
spectral capabilities below 1 keV. Assuming the {\it Chandra} observation
of Sarazin \etal to be representative, we expect that the effects of such a
component would be strongly affected by hydrogen column, but would only 
increase \Ldscr by up to a factor of two (i.e. 0.3 dex). 

One other interesting method of estimating the discrete contribution is
that of \scite{brownbreg00}. This involves fitting the surface brightness
profiles of seven elliptical galaxies, representing the hot gas
component with a King profile and the discrete sources with a de Vaucouleurs 
r$^{1/4}$ profile. Using the fitted normalisation of the de Vaucouleurs
component, Brown and Bregman find a median best fit Log L$_{dscr}$/L$_B$ =
28.51 erg s$^{-1}$, with a 99\% upper limit of Log L$_{dscr}$/L$_B$ =
29.21 erg s$^{-1}$. Although this method should in principle be able to
produce similar results to 2-component spectral fitting, these values are
somewhat lower than those found by other methods. This may be a product of
the small size of the sample, or of assuming circular symmetry to allow
1-dimensional profile fitting. It also remains to be established how well
the discrete X--ray source
population is modeled by a de Vaucouleurs profile which fits the optical
profile. The excellent spatial resolution of {\it Chandra} should allow
this question to be 
answered in the near future.   

\subsubsection{Summary}
\label{Ldscr_conc}

Possible choices for Log $L_{dscr}$ (in units of erg s$^{-1}$
L$_{B\odot}^{-1}$) are then as follows:
\begin{itemize}
\item Nearby late-type galaxies sample intercept = 29.56 $\pm$ 0.13
\item Sc galaxy sample intercept = 30.12 $\pm$ 0.06
\item Low $L_B$ sample intercept = 30.15 $\pm$ 0.11
\item \scite{ciotti91} estimate = 29.45
\item \scite{Matsushita00} hard component = 29.41
\item \scite{Blantonetal01} {\it Chandra} estimate from NGC 1553 $\simeq$
  29.52
\item \scite{brownbreg00} 99\% upper limit $\leq$ 29.21
\end{itemize}

\begin{figure*}
\centerline{\psfig{figure=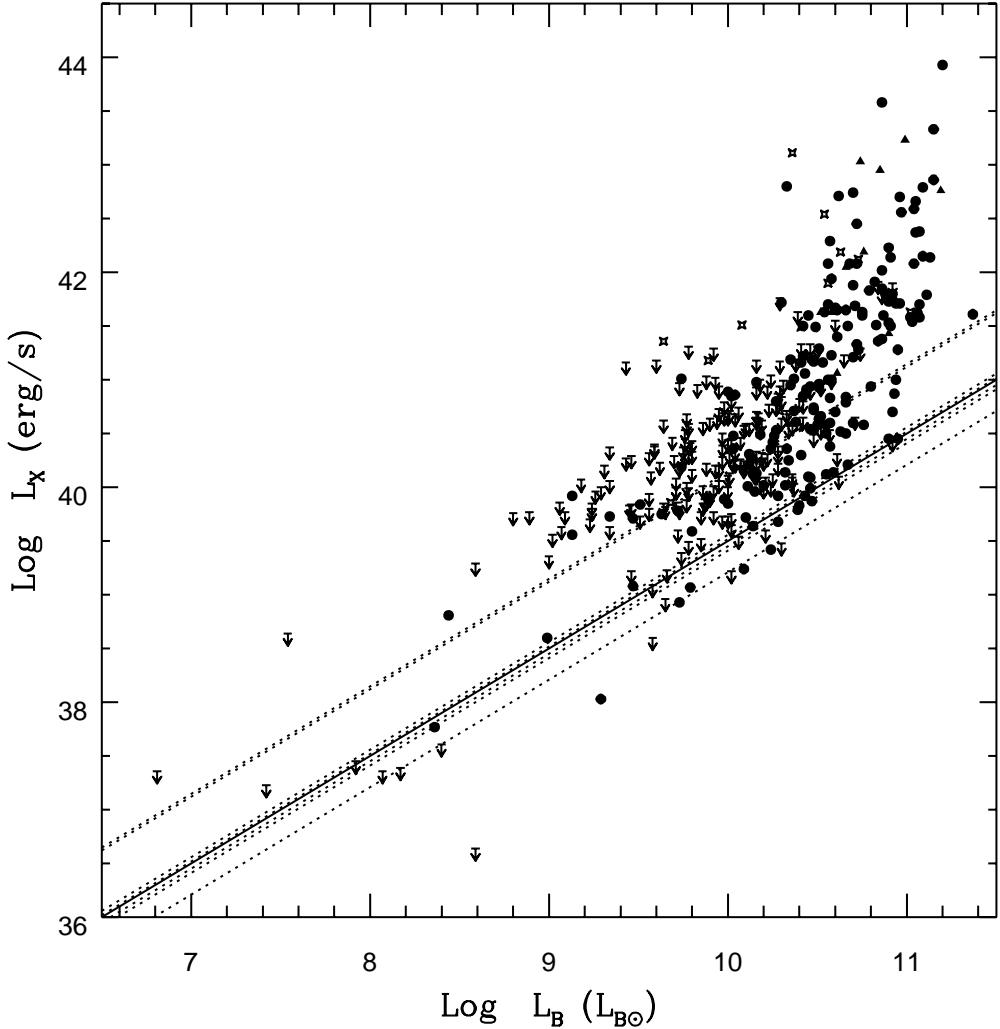,width=6in}}
\vspace{-50mm}
\caption{Plot of our early-type galaxies with \Ldscr estimates marked. The
  dotted lines represent the seven estimates listed in
  section~\ref{Ldscr_conc}, and the solid line marks \lxlb = 29.5. Point
  symbols are the same as those in Figure~\ref{2ndlxlb}.}
\label{Ldscr}
\end{figure*}

These values are compared with our data in Figure~\ref{Ldscr}.

As the two higher values are derived from samples in which $L_X$ may
include emission from gas and bright point sources, they do not seem reliable
options. The Brown \& Bregman value is considerably lower than the other
estimates, particularly as it is an upper limit. The values from Ciotti
\etal, Matsushita \etal, Blanton \etal and the nearby 
galaxies sample agree within errors, and would seem to be a reasonable
choices. A value of Log \Ldscr = 29.5 L$_{B\odot}$ 
lies between the four and is close to
being the average. This value is marked in Figure~\ref{Ldscr} as a solid
line. This value cannot be considered to be a ``hard'' lower limit; as the
plot shows, a number of our data points lie below this line. For dwarf
galaxies (L$_B <$ 9.0), we may expect to see quite large variations in
L$_X$. Each dwarf needs only a small number of LMXBs to produce the expected
luminosity of 10$^{36-38}$ erg s$^{-1}$, so minor variations in population
can produce large changes in integrated luminosity. In larger galaxies this 
statistical variation is less important, but some degree of scatter in
L$_X$ may be expected to result from factors such as different galaxy
evolutionary histories. It is also worth noting that, as discussed in
Section~\ref{Specfit}, we expect to underestimate the luminosity of
galaxies whose X--ray emission is primarily from LMXBs, owing to our
assumption of a 1 keV MEKAL model. With the 
exception of NGC 5102, all our detected non-dwarf galaxies are within a
factor of three of our \Ldscr estimate. All upper limits are also within
this range, although NGC 1375 has a  
luminosity of almost exactly \Ldscr/3. Given the factor of two expected
from underestimation of L$_X$ in these galaxies and the factor of four
variation in \Ldscr found by \scite{Kregenowetal01}, we conclude that our
data are consistent with our estimate of \Ldscr, within the expected errors.

\subsubsection{The L$_X$-L$_{dscr}$:L$_B$ relation}
\label{subtracted}
\begin{figure*}
\centerline{\psfig{figure=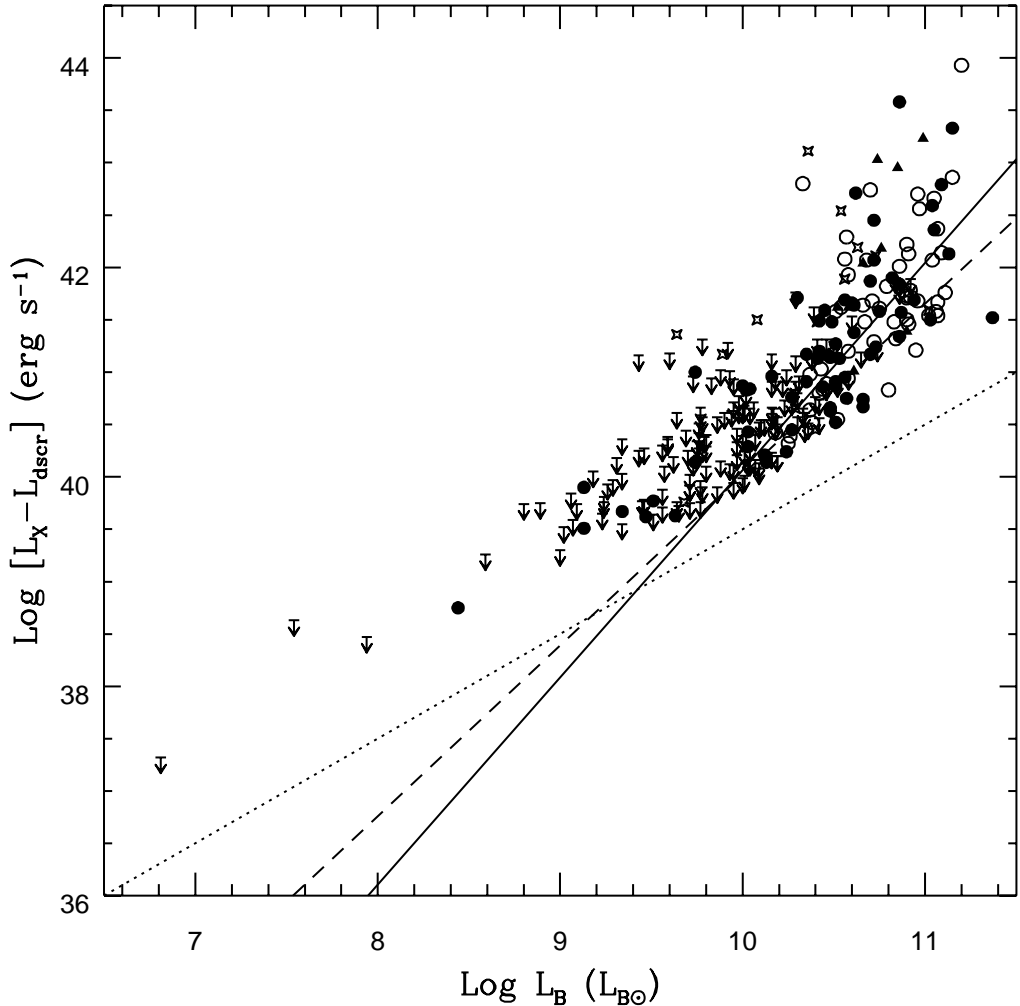,width=6in}}
\vspace{-50mm}
\caption{\label{Lx_Ldscr} Log [L$_X$-L$_{dscr}$] against L$_B$ for all
  galaxies in our main catalogue with Log L$_X$/L$_B>$30.1
  (4$\times$L$_{dscr}$). Open circles are BGGs, triangles and BCGs, stars
  are AGN, filled circles are normal detected galaxies and arrows are upper 
  limits. The Solid line is the best fit to the sample excluding AGN, BCGs
  and dwarf galaxies. The dashed line is the best fit to a very
  conservative sample which also excludes BGGs, galaxies at a distance$>$
  70 Mpc, NGC 5102 and NGC 4782. The dotted line shows our estimate of
  L$_{dscr}$=29.5.} 
\end{figure*}

Using our value of L$_{dscr}$, we can now examine how removing stellar
emission affects our \lxlbtwo relation. This should provide us with a more
accurate measure of the relation between the luminosity of the galaxies'
gaseous halos and their optical luminosity. As we expect a real variation of 
a factor $\sim$4 in \Ldscr between galaxies, we cannot simply subtract the
mean expected \Ldscr from all values of L$_X$. To do so would produce
extremely low values of L$_X$ for galaxies with total luminosities similar to
L$_{dscr}$, strongly biasing any fits. We have therefore removed all
galaxies which have  
L$_X$ values within a factor of 4 of \Ldscr, and subtracted the mean
expected \Ldscr from the
remainder. Excluding AGN, BCGs and galaxies with L$_B<$9 \LBsol, this leaves
us with a total sample of 257 points, of which 136 are upper
limits. Fitting this dataset we find slopes of 2.03$\pm$0.1 (EM) and
2.00$\pm$0.13 (BJ). The standard deviation about the regression line is
0.58 in both cases. If we further exclude galaxies to form a very conservative
sample as described in Section~\ref{lxlb_total}, the slopes of the fits
are lowered to 1.63$\pm$0.13 (EM) and 1.60$\pm$0.14 (BJ). The data and these 
fits are shown in Figure~\ref{Lx_Ldscr}.

These fits are very similar to those produced by fitting the \lxlbtwo
relation to the sample as a whole (see Table~\ref{envlxlb} for a
comparison). This is to be expected, as although 
\Ldscr is comparable to the \lxlb values of some of the galaxies, the
majority have luminosities inconsistent with X-ray emission from discrete
sources alone. As these galaxies are dominated by gas emission, subtraction 
of the discrete source contribution has little effect on their overall
luminosity. In order to test the robustness of this result, we also fitted 
samples based on excluding galaxies with X--ray luminosities within 
factors of 2 and of 6 of L$_{dscr}$. In the latter case, the slopes are slightly steeper
when excluding AGN, BCGs and dwarfs ($\sim$1.9) and slightly shallower for
the more conservative sample ($\sim$1.5). This is likely to be an effect of 
the smaller number of data points (136) and larger numbers of upper limits
(97). However, when we use a factor of 2 the slopes are less well defined;
for the sample excluding AGN, BCGs and dwarfs we find slopes of
2.14$\pm$0.11 (EM) and 2.31$\pm$0.14 (BJ). For the conservative sample, the 
slopes are again shallower, but still not in good agreement, with values of 
1.61$\pm$0.14 and 2.08$\pm$0.19. As these datasets are similar in size to
the corresponding samples used in fitting the \lxlbtwo relation, this
scatter is probably a product of the underlying distribution rather than an
effect of subtracting \Ldscr.

Figure~\ref{binnedlxlb} (Section~\ref{fit_test}) shows that at low L$_B$
the slope of the \lxlbtwo 
relation breaks and becomes shallower. A possible explanation for this is
that the observed X--ray luminosities are the result of a combination of
emission from discrete sources and hot gas. A break in the slope would then 
suggest that the shallower section is dominated by the discrete sources
while the steeper section is more influenced by gas emission. Similarly,
the lowest binned point in Figure~\ref{binnedlxlb} will have 
an X--ray luminosity dominated by discrete source emission, whereas the fit 
lines will describe the relation for gas emission. If this is the case,
then subtracting the mean value of \Ldscr expected for the lowest bin
should move  
the point downwards, into agreement with the line fits. We calculate that
the mean value of L$_X^{gas}$ for the bin is L$_X$-L$_{dscr}$=38.96, which
is in marginal agreement, at the high L$_B$ end of the bin, with the EM
fit. However, it is important to note that we expect to underestimate the
luminosities of galaxies which are dominated by discrete source emission by 
a factor of $\sim$2, due to our use of an inappropriately
soft spectral model. This may mean that the mean L$_X$ value calculated for 
the bin is also underestimated. It may also be important to take into
account the expected variations in \Ldscr between galaxies. Although we
expect a scatter of a factor of $\sim$4, some of the galaxies in this low
L$_B$ bin have very low X--ray luminosities, which could be significantly
affected by small differences in the number of X-ray binaries they contain.
We cannot therefore be certain, on the basis of the data presented here, 
that the
break in the relation is caused by the change from gas dominated to
discrete source dominated galaxies.

\subsection{Environmental Dependence of \lxlb}
\label{rhosection}
Although there are several suggested mechanisms by which the environment of a
galaxy can affect its X--ray properties, the actual role these effects play
is unclear. The observational evidence is conflicting and often difficult
to interpret.

\scite{whitesarazin91} found that for a sample of early--type galaxies
studied by {\it Einstein}, galaxies with Log \lxlb $<$ 30 (erg s$^{-1}$
L$_{B\odot}^{-1}$) had $\sim$ 50\% more neighbours than X--ray bright galaxies.
They attributed this to ram--pressure stripping, which would be expected to
reduce L$_X$ more in higher density environments. An opposite view was
presented by \scite{brownbreg99}, who found that \lxlb increased with
environmental density. Their explanation was that for the majority of
galaxies (with the possible exception of those in the densest environments)
ram--pressure stripping is a less important effect than the stifling of
galactic winds by a surrounding intra--group or --cluster medium. In this
model, the IGM/ICM encloses the galaxy, increasing the gas density of its halo
and therefore its X--ray luminosity.

Brown \& Bregman (1999) claimed an environmental dependence based
on a correlation between \lxlb and Tully
density parameter $\rho$ (\pcite{tullycat}) for their 34 galaxies. However, 
\scite{Steve} show that group--dominant galaxies, of which there are several 
in Brown \& Bregman's sample, often have X--ray luminosities which are
governed by the properties of the group rather than the galaxy. Their
high luminosities are more likely to be caused by a group cooling flow than 
by a large galaxy halo. Once these objects are removed from consideration,
the correlation between \lxlb and $\rho$ is weakened to a $\sim$1.5$\sigma$ 
effect (\scite{Steve}).

\begin{figure*}
\centerline{\psfig{figure=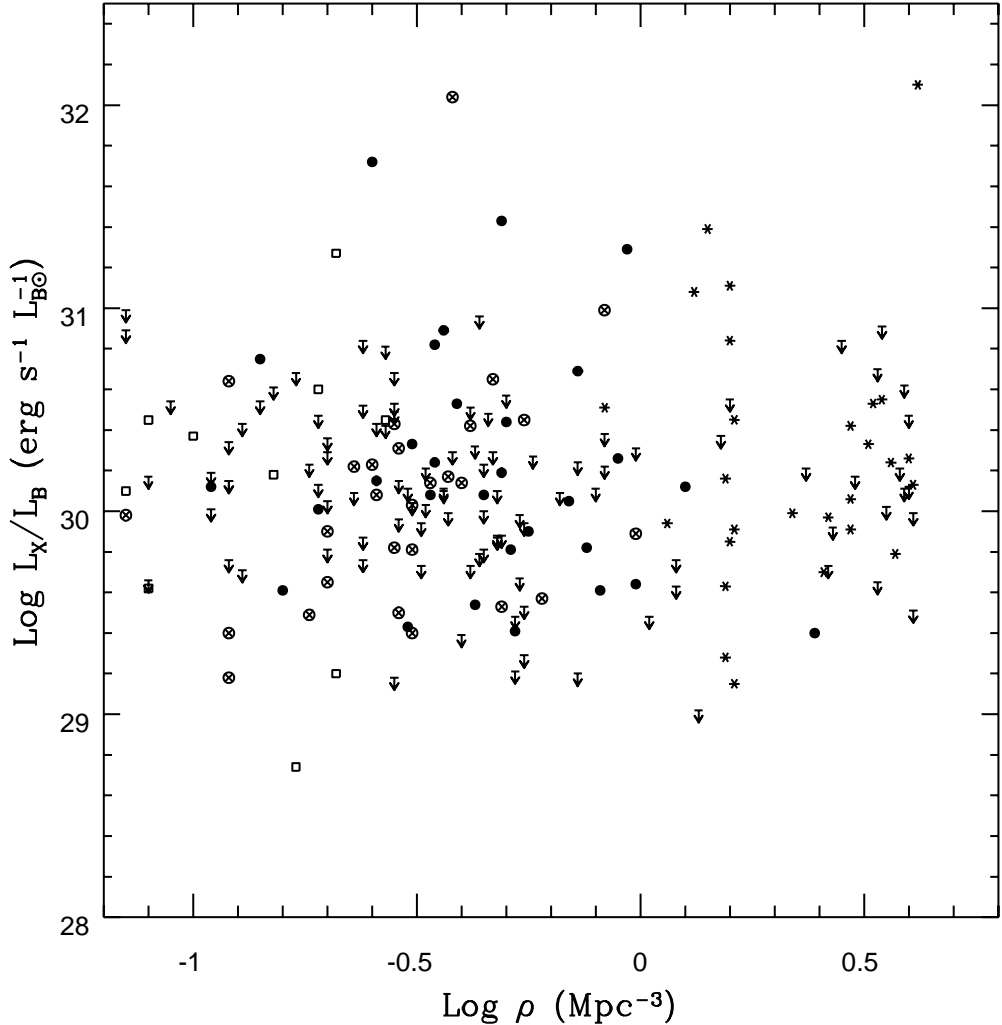,width=6in}}
\vspace{-50mm}
\caption{\label{Tullyrho} Plot of normalized X--ray luminosity against
  environmental density. Open squares are field galaxies, filled circles
  non--central group galaxies, crossed circles BGGs, asterisks cluster
  galaxies and arrows upper limits of all types. }
\end{figure*}

Our larger sample of galaxies gives us the opportunity to study this
correlation over a wide range of L$_X$, L$_B$ and environmental density. If 
Figure~\ref{Tullyrho} we therefore plot \lxlb against $\rho$ for 196 of our
galaxies listed in the Tully catalogue.

Galaxies in this plot are
subdivided into group, cluster and field samples. Cluster
membership was based on the  \scite{abellcat89} and \scite{7sam} catalogues 
while group membership was taken from \scite{Garcia93}. In
total, this gives 50 cluster galaxies and 113 group galaxies. Brightest
Group Galaxies were also taken from \scite{Garcia93}
and it is important to remember that these objects are only brightest
optically, not necessarily the dominant galaxy at the center of the group
or group X--ray
halo. However, we believe that the majority are actually group--dominant
galaxies. The group subset contains 37 BGGs. The remaining 33 galaxies not
listed in the cluster or group catalogues were assumed to lie in the
field. This is probably the weakest classification and is likely to be
contaminated to some extent with galaxies at the edges of clusters and groups.

The plot shows no obvious trend, but to check for weak correlations we used 
the statistical tests described in section~\ref{surv}. No trend was found
in the sample as a whole, nor in any of the subsamples. We also calculated
mean \lxlb values for each of the subsamples, excluding all AGN, BCGs
and dwarf galaxies. These values are shown in
table~\ref{envmean}. The field, group and cluster subsamples have similar
mean values, while the BGG subsample has a slightly larger mean \lxlb,
 as might be expected from the previous results.

\begin{table}
\begin{center}
\begin{tabular}{lcc}
Subset & mean \lxlb & Error\\
\hline
Cluster & 29.733$^*$ & $\pm$0.094 \\
Field & 29.548 & $\pm$0.196 \\
Group (total) & 29.908 & $\pm$0.066 \\
Group (non-BGG) & 29.719$^*$ & $\pm$0.065 \\
Group (BGG) & 29.977 & $\pm$0.096 \\
\end{tabular}
\end{center}
\caption{\label{envmean} Mean \lxlb values for the environmental
  subsamples shown in Figure~\ref{Tullyrho}. Values marked by an asterisk
  may be slightly biased as the lowest value in the sample was an upper limit.}
\end{table}

\begin{figure*}
\centerline{\psfig{figure=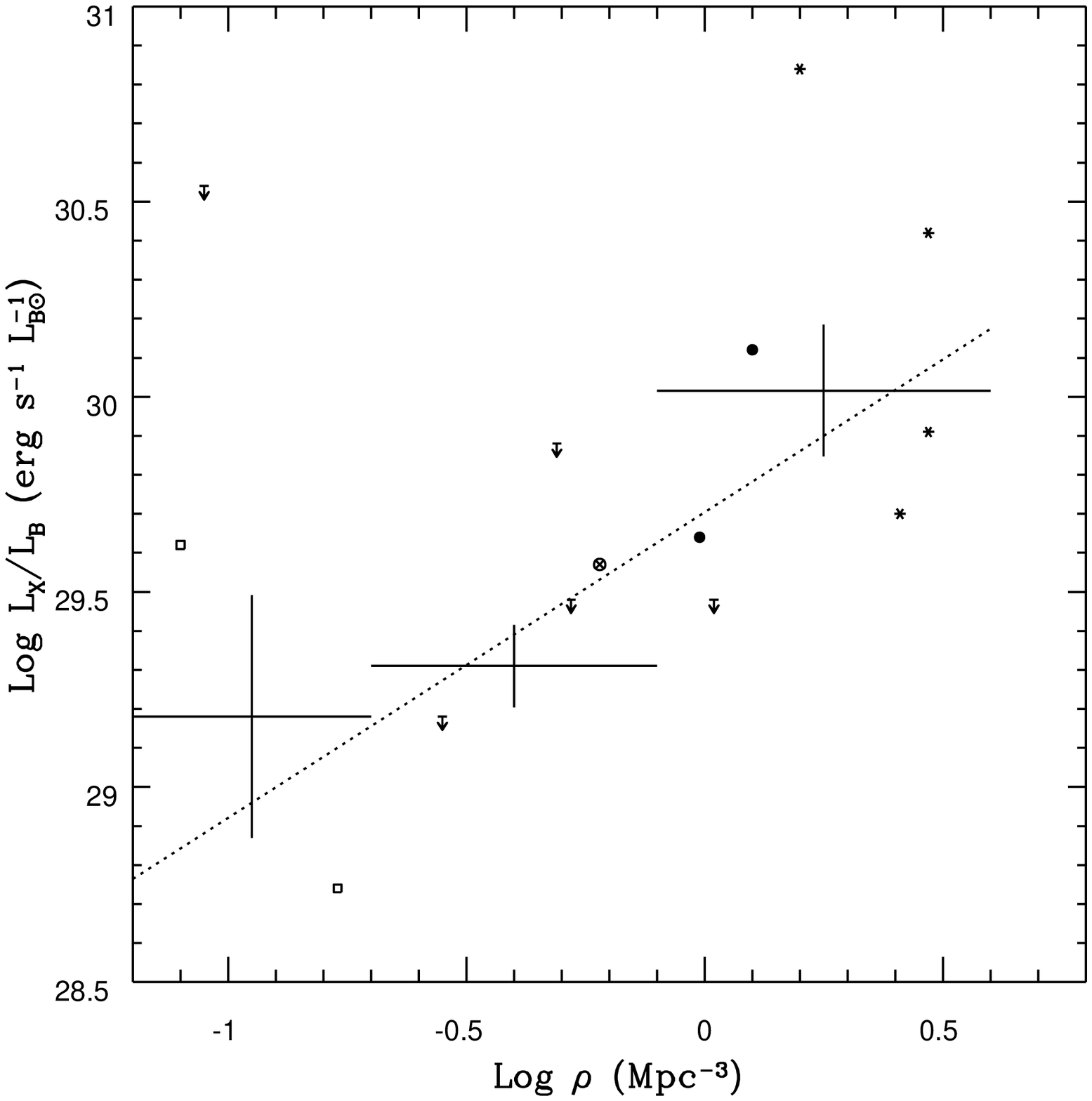,width=6in}}
\vspace{-50mm}
\caption{\label{BBripoff} Plot of \lxlb against $\rho$ for non--BGG galaxies 
  from the sample of Brown \& Bregman (1999). \lxlb values are from our
  catalogue, $rho$ values are from Tully (1988). The dotted line is the
  best fit to the data, and the large crosses are mean values for the three
  bins described in the text. Other symbols are the same as in Figure~\ref{Tullyrho}.}
\end{figure*}

The lack of a general correlation is surprising, as the previous studies suggest we
should find at least a weak trend. As we are using the same method as Brown
\& Bregman, we decided to check for a correlation in their sample of
galaxies using our own X--ray luminosities. These data (excluding galaxies
identified as BGGs in Helsdon \etal) are plotted in Figure~\ref{BBripoff}.

Although there is more scatter in \lxlb than seen in Brown \& Bregman's
plot, a trend for increasing \lxlb with environmental density is
clear. Statistical tests show the correlation to be at least 97\%
(2.5$\sigma$) 
significant, with a best fit slope of 0.78 (EM) or 0.75 (BJ). Binning the
data in the same ranges as used by Brown \& Bregman and Helsdon \etal
produces the three large crosses in the plot. These also clearly show a
trend, despite the fact that the centre and right hand crosses are likely
to be biased downwards as the lowest points in these bins are upper
limits. 

As the trend is seen in this sample of galaxies but not in our more general 
catalogue, it seems likely that it is a product of the sample selection
process. Brown \& Bregman's sample is composed of the 34 optically
brightest galaxies chosen from \scite{Faberetal89}, excluding AGN and dwarf 
galaxies. The selection of bright galaxies has one clear effect; more than
half of their galaxies are BGGs, likely to have unusually high X--ray
luminosities. Of the remaining galaxies, three are found in the field, six
are in groups and five in clusters. Of the cluster galaxies, the two most
X--ray luminous are NGC 1404, a large E1 galaxy in the Fornax cluster, and
NGC 4552 (M89), one of the large ellipticals in the Virgo cluster. The high
L$_X$ values of these two galaxies and the low L$_X$ value
of NGC 5102 have a strong influence on the fitted slope. Indeed, when using
our data, their removal eliminates the correlation altogether.
NGC 4552 
is known to be a Seyfert 2 (\pcite{Veron}), and therefore its X--ray
luminosity is likely to be misleading. NGC 1404 may also be an unusual
case, as it lies within the X--ray envelope of NGC 1399 and may
be interacting with it (\pcite{Forbesetal98}). As mentioned in
Section~\ref{Res2}, NGC 5102 is a recent ($\sim$400 Myr) post--starburst
galaxy and has a population of young blue stars (\pcite{BicaAlloin87})
which may have `artificially' raised its B band luminosity.  

\subsection{The \lxlbtwo relation in different Environments}
\label{FGCsection}
In order to gain another viewpoint on the relation between environment and
X--ray luminosity, we decided to examine the \lxlbtwo relation in field, group
and cluster environments. We split the sample as described above, but no
longer limited ourselves to galaxies listed in the Tully catalogue. The
sample therefore included all galaxies 
within 5,500 km s$^{-1}$ (the limit of the Garcia (1993) group catalogue). The
resultant fits are shown in Table~\ref{envlxlb}, along with the fits to the 
field, group and cluster sets. In each case we have removed cluster central 
galaxies, AGN, dwarf galaxies and galaxies at distances $>$70 Mpc before
fitting. The cluster sample 
contains 57 objects (of which 36 are upper limits), the field sample 76
objects (55 upper limits) and the group sample 185 objects (85 upper
limits). Separating BGGs from the group sample gives a non--BGG sample of
116 objects (69 upper limits) and a BGG sample of 69 objects (16 upper limits).

\begin{table}
\begin{center}
\begin{tabular}{lcccc}
Subset & Fit & Slope (Error) & Intercept (Error) \\
\hline
Combined & EM & 1.63 ($\pm$0.14) & 23.38 ($\pm$1.41)\\
 & BJ & 1.94 ($\pm$0.17) & 20.13 \\
L$_X$-L$_{dscr}$ & EM & 1.63 ($\pm$0.13) & 23.70 ($\pm$1.36)\\
 & BJ & 1.60 ($\pm$0.14) & 24.11\\
Cluster & EM & 1.77 ($\pm$0.27) & 21.91 ($\pm$2.71)\\
 & BJ & 1.77 ($\pm$0.29) & 21.93\\
Field & EM & 1.62 ($\pm$0.23) & 23.58 ($\pm$2.40)\\
 & BJ & 1.61 ($\pm$0.26) & 23.61\\
Group (total) & EM & 1.92 ($\pm$0.13) & 20.56 ($\pm$1.39)\\
 & BJ & 1.90 ($\pm$0.16) & 20.73\\
Group (non--BGG) & EM & 1.62 ($\pm$0.14) & 23.57 ($\pm$1.39)\\
 & BJ & 1.59 ($\pm$0.16) & 23.85\\
Group (BGG) & EM & 2.58 ($\pm$0.36) & 13.60 ($\pm$3.76)\\
 & BJ & 2.57 ($\pm$0.40) & 13.71\\
\end{tabular}
\end{center}
\caption{\label{envlxlb} Best fit \lxlbtwo lines for galaxies in field, group
  and cluster environments. All subsets exclude AGN, BCGs, dwarf galaxies
  and galaxies at distances $>$ 70 Mpc. The Combined values for the
  complete sample {\em excluding BGGs}, and the best fits to the
  L$_X$-L$_{dscr}$:L$_B$ relation are shown for comparison.}
\end{table}

\begin{figure*}
\parbox[t]{1.0\textwidth}{\vspace{-1em}\includegraphics[width=9.5cm]{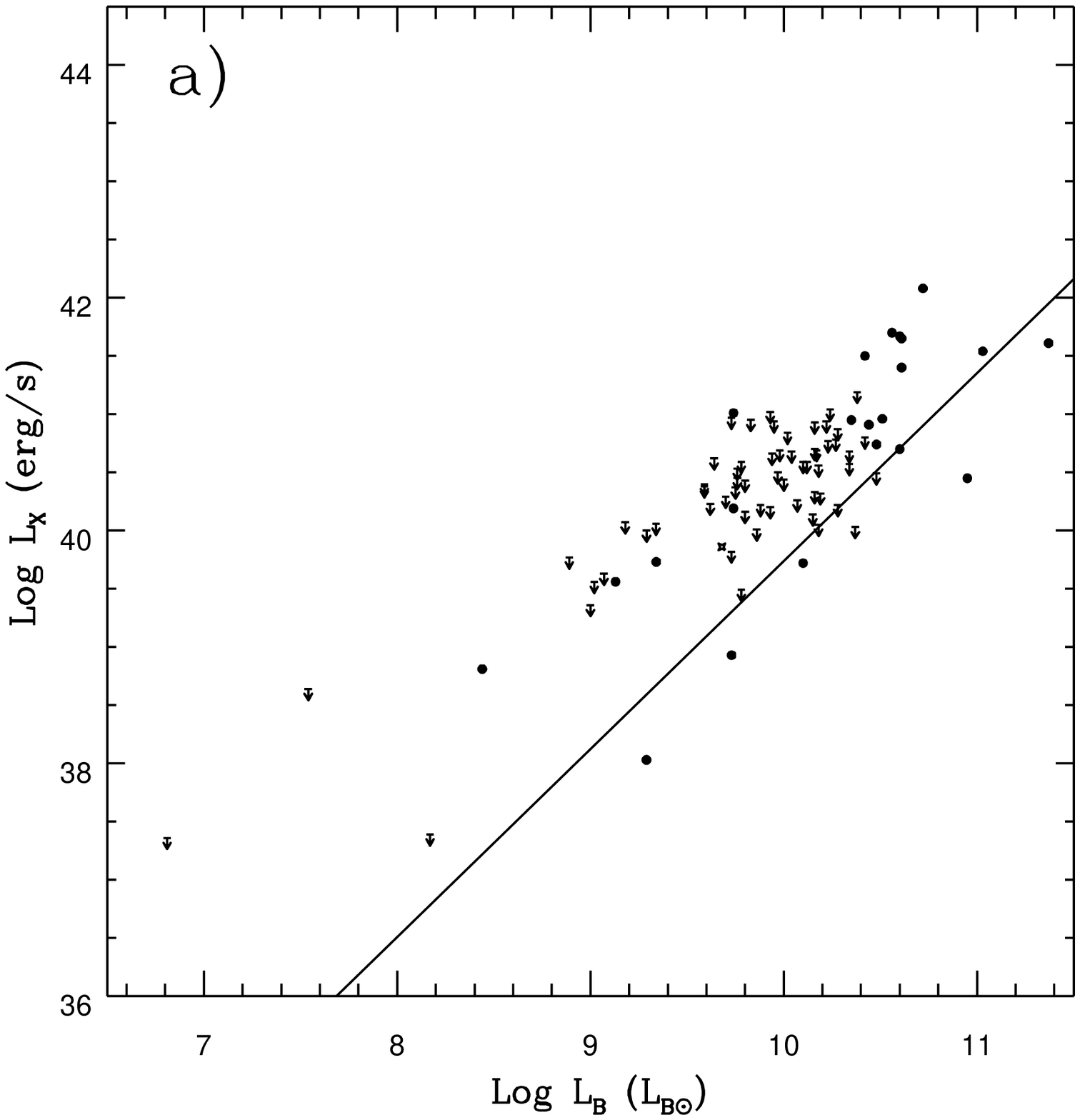}}
\parbox[t]{0.0\textwidth}{\vspace{-12.7cm}\includegraphics[width=9.5cm]{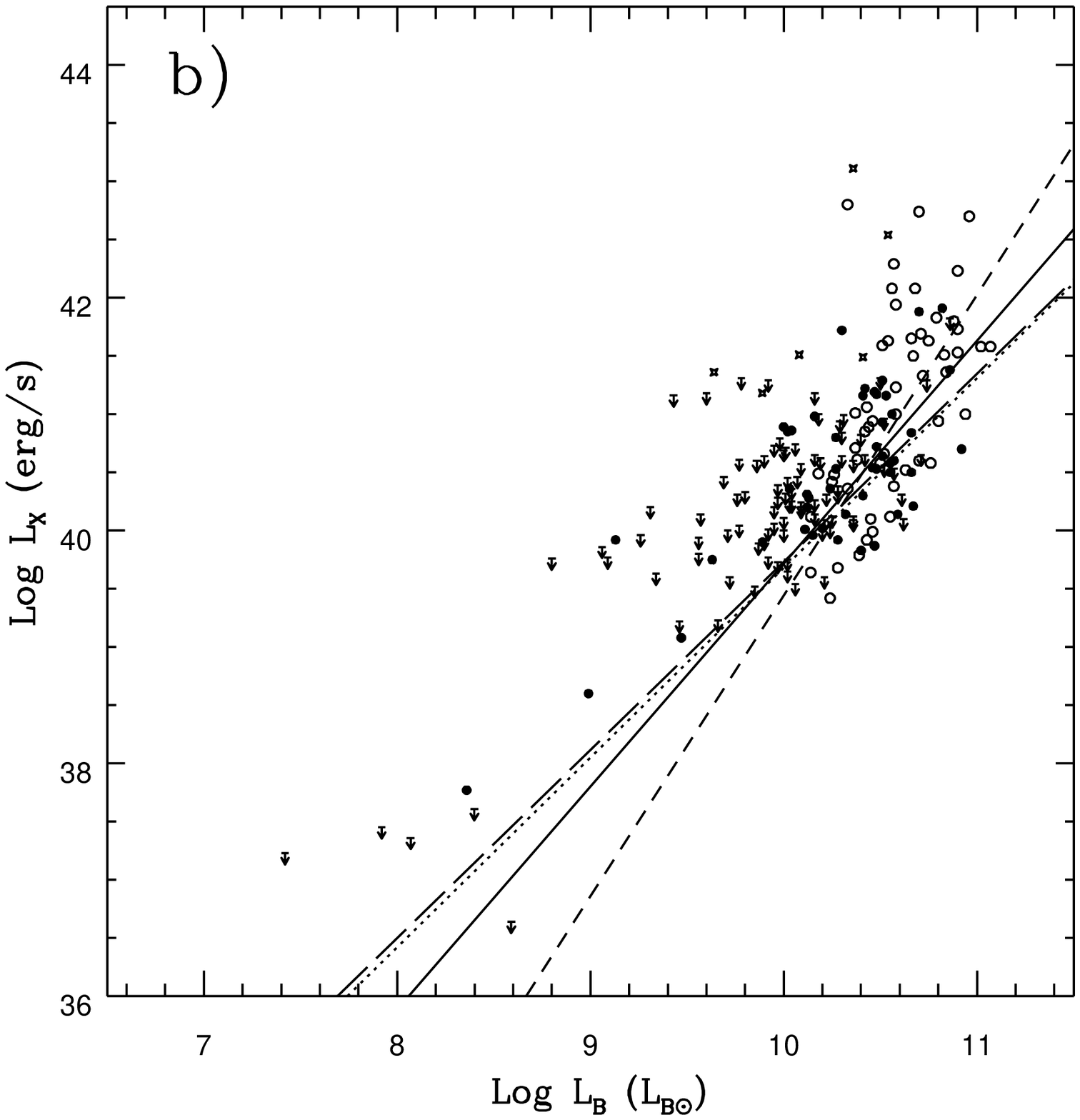}}
\parbox[t]{1.0\textwidth}{\vspace{-4cm}\includegraphics[width=9.5cm]{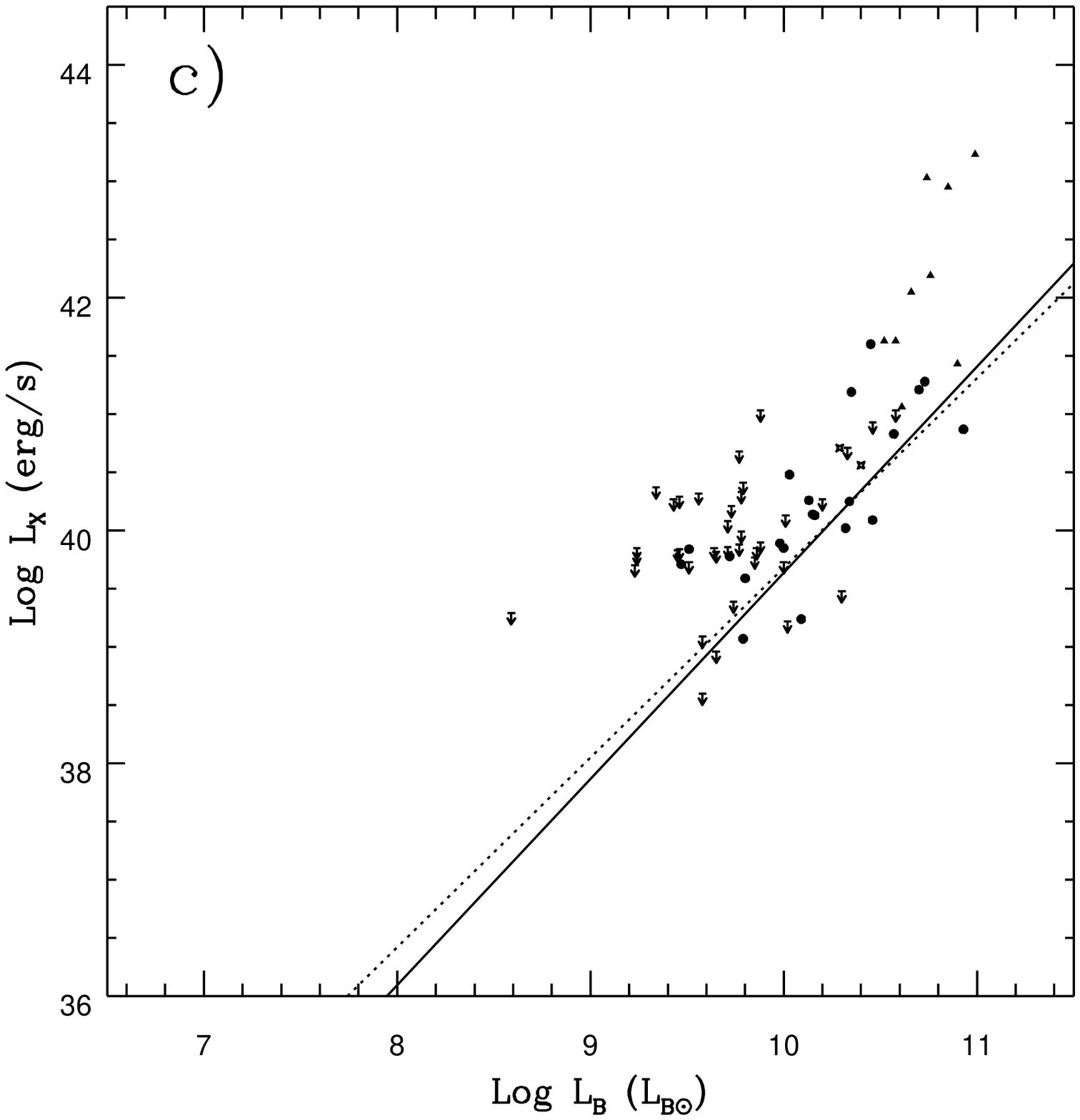}}
\vspace{-3cm}
\hspace{8.5cm}
\begin{minipage}{7cm}
\vspace{-19.7cm}
\caption{\label{FGClxlb} \lxlb data for a) field, b) group and c) cluster
  subsets. In each plot the solid line represents the best fit to the
  sample, excluding AGN (marked as stars), BCGs (marked as triangles),
  dwarfs and galaxies at distances $>$70 Mpc. In Fig. b, open circles
  represent BGGs, and the two dashed lines show fits to the BGGs (long
  dashes) and to the remaining sample with BGGs excluded (short
  dashes). The dotted line in Figures b and c represents the EM fit to the
  complete catalogue with exclusions as above. This fit is not shown in
  Figure a, as it is indistinguishable from the fit to the subsample.} 
\end{minipage}
\end{figure*}

Figure~\ref{FGClxlb} shows plots 
of the cluster, field and group data with best fit lines. In terms of
L$_B$, it is notable that although the field and group sets cover a
similar range, there are very few optically faint cluster galaxies. This is 
likely to be caused by the difficulty of observing small, X--ray faint
galaxies in an X--ray bright ICM. In the field no such problem 
occurs, and many groups are faint enough to allow such small objects to be
observed. Both group and cluster data show a number of highly X--ray
luminous objects, probably giant ellipticals and group or cluster dominant
galaxies at the centers of large X--ray halos. In the field, only one galaxy 
(NGC 6482) has Log L$_X>$ 42 erg s$^{-1}$ and the high end of the \lxlbtwo
line is sparsely populated. Comparing \lxlbtwo slopes shows a similar
trend, with the BGG sample producing the steepest slope, then cluster
galaxies, non--BGGs and lastly field galaxies with the shallowest
relation. The slope of the BGG sample is similar, within errors,  to that
of the best fit line to the Brown \& Bregman sample. 

It is interesting to note that the slopes for the field, cluster and
non-BGG group samples are all similar, within errors. Table~\ref{envlxlb}
also shows the fits to the catalogue as a whole, excluding AGN, BCGs, BGGs, 
dwarfs and galaxies at distances $>$ 70 Mpc. The EM fit to this supersample 
agrees with the fits to the three subsamples, and the BJ fit has overlapping 
errors with the field and cluster subsamples. This suggests that the
\lxlbtwo relation may be similar for the different environmental subsets
when biasing objects are excluded. 
To further investigate this similarity, we have fitted fixed
slope lines to the field, group, cluster and combined subsets described
above. As the EM and BJ algorithms find slopes of 1.63 and 1.94 for the
combined subset, we use these values. Table~\ref{envint} shows the resulting
intercepts and errors, calculated using the Kaplan-Meier estimator. In both
cases, the intercepts for 
the field, cluster and non-BGG group subsets agree within errors, and also
agree with the intercept of the combined subset, whilst the BGG 
intercept is markedly higher. Our results suggest, then, that with the 
exception of BGGs, early-type galaxies have a universal 
mean \lxlbtwo relation which is unaffected by environment. 

\begin{table}
\begin{center}
\begin{tabular}{lcc}
 & Slope = 1.63 & Slope = 1.94 \\
Subset & Intercept (Error) & Intercept (Error) \\
\hline
Combined & 23.446 ($\pm$0.048) & 20.252 ($\pm$0.054) \\
Cluster & 23.384 ($\pm$0.091) & 20.253 ($\pm$0.096) \\
Field & 23.446 ($\pm$0.109) & 20.247 ($\pm$0.126) \\
Group (total) & 23.457 ($\pm$0.049) & 20.323 ($\pm$0.051) \\
Group (non-BGG) & 23.473 ($\pm$0.062) & 20.266 ($\pm$0.069) \\
Group (BGG) & 23.667 ($\pm$0.083) & 20.401 ($\pm$0.081) \\
\end{tabular}
\end{center}
\caption{\label{envint} Best fit intercepts to \lxlbtwo relations with
  fixed slopes as shown in the table. All values are calculated using the
  Kaplan-Meier estimator. All subsamples exclude AGN, BCGs, dwarf galaxies
  and galaxies at distances $>$ 70 Mpc. The Combined subset also excludes BGGs.}
\end{table}

\section{Discussion and Conclusions}
\label{discuss}
\subsection{The \lxlbtwo Relation for Early--type Galaxies}
The slope of the \lxlbtwo relation has been the subject of debate for some
time. As an indicator of how gas properties change with galaxy mass
(and therefore probably with time) it is an important relation to
measure precisely. The difficulties associated with accurately measuring a
large sample of galaxies and avoiding contamination from other X--ray
sources has made this difficult. Our sample has some advantages; we
are able to remove some group/cluster contamination from our luminosities,
and we have a large enough sample to allow us to exclude problem galaxies
without making fitting impossible. We are also able to remove BGGs as well
as BCGs, which
in principle allows us to define a sample of galaxies unbiased by emission
from cluster-- or group--scale cooling flows. 

Our initial fits agree fairly well with previous estimates of the
\lxlbtwo relation (\pcite{Beuingetal99}; \pcite{donnellyfaberoconnell90};
\pcite{whitesarazin91}), providing evidence that our catalogue is similar
to the 
samples used for those fits. This is to be expected, as our sample was
selected in a similar way, and contains galaxies (and in some cases data)
in common with these fits. However, the line fits for our sample
excluding BGGs are somewhat shallower, particularly if we adopt the
precaution of removing galaxies whose distance makes their surroundings
uncertain. This change in slope shows that the BGGs are, as predicted,
steepening our fits. The implication of this result is that previous
\lxlbtwo values have also been influenced by the inclusion of BGGs (and
possibly BCGs, AGN, etc). 

As our data are drawn from samples of galaxies which have been observed and
analysed in different ways, we have considered potential problems arising
from their combination. The basic \lxlbtwo relation for the {\it ROSAT}
data we have analysed agrees fairly well with that of
\scite{fabbianokimtrinchieri92}, but less well with the relation of
\scite{Beuingetal99}. This appears to be caused by the difference in data
quality and analysis strategies employed. In particular, the low numbers of 
counts in RASS images can make separation of close pairs of sources
difficult, and also makes it difficult to distinguish between emission
associated with a cluster or group and its central galaxy. These
difficulties have lead to overestimation of some galaxy luminosities in
Beuing \etal. The corrections described in Section~\ref{catconv} should
counteract this effect to some extent, but we cannot expect all data points to
be completely accurate.
  
Testing our methods of line fitting suggests that of the available
algorithms we are probably using the most appropriate. However, the
\lxlbtwo relation does not appear to be well described by a single powerlaw 
fit, even when we remove objects whose luminosities are likely to be
dominated by AGN, cooling flows or stellar emission. It seems more likely
that it is the product of the combination of discrete source emission, with an 
\lxlbtwo slope of approximately unity, and gas emission with a steeper
slope. For 
the catalogue as a whole, a third, even steeper component is added by the
cooling flow enhanced luminosities of group and cluster dominant
galaxies. This combination of emission mechanisms may explain the
variety of slopes which have been measured in previous studies; the slope
will be dependent on the range of L$_B$ used and the number of group/cluster
dominant galaxies included. However, the small number of detected
galaxies and large expected variations in L$_X$ at L$_B\simeq$9 make this
model difficult to test conclusively with our data.

\subsection{Environmental Dependence of \lxlbtwo}
The results presented in Sections~\ref{rhosection} and~\ref{FGCsection}
lead us to three main conclusions: 
\begin{itemize}
\item There is no strong correlation between \lxlb and environmental
  density ($\rho$).
\item BGGs have a significantly steeper \lxlbtwo relation that non-BGG
  group galaxies.
\item Once objects such as BGGs, BCGs, AGN and dwarf galaxies are excluded, 
  the \lxlbtwo relation is largely independent of environment. 
\end{itemize}

When considering the first of these points it is important to note that
the lack of a trend with $\rho$ does not mean that environment has no effect
of the galaxy X--ray properties. There are numerous suggestions of
processes which can affect the X--ray halo of a galaxy. Ram-pressure
stripping is likely to remove gas from galaxies passing through a dense
intra--cluster medium (\pcite{GunnGott72}), and turbulent viscous stripping 
may be as effective in the group environment (\pcite{Nulsen82}). It is also 
likely that the IGM and ICM provide reservoirs of gas which can be captured 
by slow moving or stationary galaxies. Stifling of galactic winds
(\pcite{brownbreg99}) may also play an important role in increasing the
X--ray luminosity of some galaxies. Our results do not rule out these
processes.

The lack of trend does however suggest that, in most environments, none of
the processes affecting  
X--ray halos is dominant. It is possible that the processes interact and
counter--balance one another, or that the mechanisms are less efficient
than thought and only affect galaxies in the very densest environments.
It is also probable that the interactions between
group/cluster and galaxy halos are more complex than we have
assumed. Observations of one $z \sim 0.83$ cluster (\pcite{vandokkum99})
have provided evidence to support the theory that clusters are products of
the the merger of previously formed groups of galaxies. The evidence of
sub-clumping within local clusters suggests that the halos of these
groups can survive the merger process. In this case, ram--pressure
stripping of gas from the galaxies within the groups seems
unlikely. On the other hand, modeling studies strongly suggest that a
field galaxy falling into the dense core of a relaxed cluster is very
likely to be ram--pressure stripped of the majority of its gas (\eg
\pcite{Quillisetal00}). Whether a galaxy is likely to have been stripped
depends not only on its position, but on the state of the cluster when that 
galaxy entered it, whether it fell in as part of a group and how much of the 
group halo survived, and probably on many other criteria. A comparison of
\lxlb with $\rho$ may well be too simple a test to tell us much more about the
interactions which take place. It is worth remembering that $\rho$ is a
measure of the local density of galaxies, whereas most of the mechanisms
mentioned above depend on the gas density encountered by a galaxy over the
past few gigayears.

The results of the \lxlbtwo fits for galaxies in different environments
also suggest that although environment may affect individual galaxies, it
cannot change the nature of whole populations. The fact that the \lxlbtwo
relation is similar in field, group and cluster environments provides 
strong evidence that X--ray halos are not radically different in these
different environments. It seems more likely that the X--ray properties of
early--type galaxies are governed by internal processes, with outside
influences in most cases
producing scatter rather than a complete change. This fits well with the
results of our previous paper (\pcite{OFP00age}) in which we showed a trend
in \lxlb with galaxy age. This trend appears to be driven by the evolution
of galaxy winds which produce a general increase in the size and density of 
the X--ray halo as the galaxy ages after a major starburst. Galaxy wind
models predict that the amount of gas produced and retained by a galaxy
depends on its mass and on the way in which supernova heating changes with
time. In general they predict that larger galaxies will have larger halos,
but as noted in \scite{Steve}, most models have been designed to fit the
assumption that the \lxlbtwo relation is steep (L$_X \propto$ L$_B^2$). It
would be interesting to see what changes in the models are needed to
reproduce the flatter relations observed in this work.

Comparison between the slopes found in this work, and those found by
\scite{Steve} show some intriguing differences. Helsdon \etal, working with 
a sample of 33 X--ray bright groups, examined the X--ray properties of the
galaxies in those groups. The main result of the study was the strong
dissimilarity between group--dominant galaxies and all other
early--types. A second important result was that the L$_X$:L$_B$ relation for
the non-dominant galaxies appeared to have a slope of $\sim$1. In
contrast, we find that 
our cleaned sample of non-BGG group galaxies has a slope of $\sim$1.6. There
are a number of reasons why we might expect such a difference between the
two studies. (i) Our sample 
of non-BGG group is drawn from $\sim$90 groups, for many of which we only
have data on one member. (ii) These groups are optically rather than X--ray
selected, and cover a wide range of sizes. (iii) The two samples cover somewhat
different ranges in L$_B$; $\sim$9.8--11.4 for Helsdon \etal, $\sim$9--10.8
in this work. (iv) Our X--ray analysis attempts to
remove at least a part of any contamination from a surrounding group halo,
but is crude compared to the techniques used in the analysis of Helsdon
\etal. It seems safe to say that the sample of Helsdon \etal covers a
narrower range of group properties than ours, but provides a more accurate
and in--depth view of X--ray bright groups. 

Explaining the difference is
difficult. One possibility is that the Helsdon \etal sample, selected as a
sample of X--ray bright groups, represents a high range of gas densities
which our sample does not thoroughly cover. In this case, most early--type
galaxies in the groups could have suffered ram--pressure or viscous
stripping, leaving them with minimal amounts of hot gas and producing the
slope of unity. However, Helsdon \etal also found that the level of the
\lxlbtwo relation for group galaxies was a factor of $\sim$2.5 higher than
their estimate of discrete source emission. Their estimate of \Ldscr is
consistent with ours, and we must conclude that galaxies in their sample
are not entirely devoid of X--ray emitting gas. Another possibility is
that the lack of gas in these objects could be caused by the stripping of
their dark matter halos as they entered the group. In this case their lack
of mass would make retention of gas difficult. In both cases we must assume
that the sample presented in this work does not contain enough galaxies
from X--ray bright, massive, relaxed groups for the
effects of stripping to be clearly observed.

A similar problem occurs in our identification of BGGs. The Helsdon \etal
sample selects central dominant galaxies based on their position at the
centre of the X--ray halo. This ensures that the galaxy is at the centre of 
the group potential well, and so the excess emission observed is likely to
be produced by a group cooling flow centred on the galaxy. In our sample we
assume that the galaxy which is optically brightest is central and
dominant, neither of which is necessarily true. Some of the 
groups in our sample are X--ray faint, in which case any cooling flow
is likely to have a low mass deposition rate and therefore a minimal effect
on the galaxy at its  
centre. However, the steepness of the \lxlbtwo slope for BGGs indicates
how different these galaxies are from the general population. It 
is apparent that these objects can significantly bias studies which include 
them, and that to consider them simply as elliptical galaxies like any
other may be misleading.

What these considerations make clear is that group galaxies, which constitute
the majority, both in our sample and the universe as a whole, require further 
study before we can understand the processes which shape
them. We need to know if the
dominant galaxies of X--ray faint groups are as different from their
neighbours as those in X--ray bright groups. We also need to know how the
slope of the \lxlbtwo relation changes with group mass and gas density, in
order to be able to determine how the slope of unity observed by Helsdon
\etal is produced. A detailed study of a wide range of groups appears to be 
the necessary to answer these questions.

\subsection{The Discrete Source Contribution}
As the collecting area and spectral resolution of X--ray observatories has
improved, it has become more important to be able to separate the
contribution of discrete sources from that of hot gas. Late--type galaxies
are dominated in the X--ray by discrete sources, but as shown in
Section~\ref{morphsec}, there seems to be a significant hot gas component
in earlier--type spirals. Similarly in ellipticals and S0s, the brighter
galaxies are dominated by gas emission, but accurate spectral fitting
requires a discrete source component.

We have attempted to define the level of this contribution. Fitting lines
to morphological or luminosity defined spiral subsamples seems
to produce normalisations which are too high. This may be because of the
inclusion of emission from other X--ray sources such as hot gas, HMXBs,
young supernovae or short lived bright transients in some
galaxies. The result of fitting a sample for which some of the emission
from these sources has been removed is a lower normalisation, which is in
closer agreement with the lower boundary of our catalogue of X--ray
luminosities. 

Ideally we would wish to derive an average \Ldscr for early--type galaxies
from those galaxies themselves. At present, there are few data sets of
sufficient quality to allow accurate spectral fitting of a hard
component. The estimate of \scite{Matsushita00} is probably the best of
these and is also in fairly close agreement with our catalogue lower
boundary. However, {\it XMM} and {\it Chandra} will be necessary to fix
the discrete source contribution more precisely, and allow us to examine
the evolution of the gas component more precisely. In the mean time, we
believe our value of \Ldscr = 29.5 erg s$^{-1}$ L$_{B\odot}^{-1}$ to be a reasonable estimate of the mean
contribution.

\noindent{\bf Acknowledgements}\\
The authors would like to thank S. Helsdon for useful discussions and his
extensive help with {\sc asterix} software, A. Sanderson, R. Brown and
B. Fairley for their help with various aspects of this work, C. Jones for
useful comments on the paper and R. Kraft for providing unpublished
information on {\it Chandra} observations of discrete sources. We would
also like to thank the referee, J. Bregman, for several comments and
suggestions which have improved the paper.
This research has made use of the NASA/IPAC Extragalactic Database (NED)
which is operated by the Jet Propulsion Laboratory, California Institute of
Technology, under contract with the National Aeronautics and Space
Administration. This research has also made use of data obtained from the
Leicester Database and Archive Service at the Department of Physics and
Astronomy, Leicester University, UK, and of the LEDA database
(http://leda.univ-lyon1.fr). The work made use of Starlink 
facilities at Birmingham and E.O'S. acknowledges the receipt of a PPARC
studentship.  

\nocite{robertshogg}
\nocite{fabbianokimtrinchieri92}
\nocite{Beuingetal99}
\nocite{pands96}
\bibliographystyle{mnras}
\bibliography{../paper}

\onecolumn
\begin{longtable}[c]{lcccccc}
Name & D & Log L$_B$ & & Log L$_X$ & Source & T \\
 & (Mpc) & (L$_{B\odot}$) & & (erg s$^{-1}$) & & \\
\hline
\hline
\endhead
\endfoot
ESO101--14       &  30.12 &  9.93*& $<$ & 41.02 & B & -3.0 \\
ESO107--4        &  38.89 & 10.22 & $<$ & 40.94 & B & -4.0 \\
ESO137--6        &  69.75 & 10.56 &   & 42.08 & N & -4.8 \\
ESO137--8        &  47.95 & 10.42*&   & 41.22 & N & -3.9 \\
ESO137--10       &  42.27 & 10.46*&   & 40.94 & N & -3.0 \\
ESO138--5        &  35.39 & 10.16*& $<$ & 41.18 & B & -3.0 \\
ESO148--17       &  38.36 & 10.04 & $<$ & 40.68 & B & -4.8 \\
ESO183--30       &  33.59 & 10.18 & $<$ & 41.00 & B & -3.2 \\
ESO185--54       &  56.36 & 10.84*&   & 41.36 & N & -4.8 \\
ESO208--21       &  10.36 &  9.34 &   & 39.73 & B & -3.1 \\
ESO243--45       & 100.91 & 10.84*& $<$ & 41.91 & N & -3.0 \\
ESO273--2        &   3.20 &  7.54 & $<$ & 38.64 & B & -3.2 \\
ESO286--50       &  33.31 &  9.76 & $<$ & 40.53 & B & -3.2 \\
ESO306--17       & 139.95 & 11.15*&   & 43.33 & B & -3.9 \\
ESO322--60       &  32.85 &  9.86*& $<$ & 40.60 & B & -2.1 \\
ESO351--30       &   1.99 &  8.59 & $<$ & 36.64 & N & -4.8 \\
ESO356--4        &   0.63 &  8.17 & $<$ & 37.39 & B & -4.8 \\
ESO381--29       &  32.65 &  9.78 & $<$ & 40.59 & B & -3.8 \\
ESO400--30       &  30.17 &  9.76 & $<$ & 40.45 & B & -4.0 \\
ESO425--19       &  89.40 & 10.75*&   & 41.60 & B & -3.0 \\
ESO428--11       &  10.49 &  9.07 & $<$ & 39.63 & B & -2.9 \\
ESO443--24       &  65.97 & 10.67 &   & 41.50 & N & -3.2 \\
ESO495--21       &   9.16 &  9.13*&   & 39.56 & N & -2.6 \\
ESO507--21       &  40.23 & 10.51*&   & 40.93 & B & -2.8 \\
ESO552--20       & 123.49 & 11.04*&   & 42.59 & B & -3.9 \\
ESO553--2        &  61.88 & 10.42 &   & 41.50 & B & -2.2 \\
ESO565--30       & 132.99 & 11.05*&   & 42.37 & B & -3.1 \\
E1090221        &  37.43 & 10.46 & $<$ & 40.52 & B &  0.0 \\
E920130         &  19.79 &  9.70 & $<$ & 40.29 & B & -3.8 \\
IC310           &  63.39 & 10.54 &   & 42.54 & B & -2.0 \\
IC989           & 101.33 & 10.60 & $<$ & 41.55 & F & -4.9 \\
IC1024          &  21.68 &  9.31*& $<$ & 40.20 & F & -2.0 \\
IC1459          &  18.88 & 10.37 &   & 40.71 & N & -4.7 \\
IC1531          & 100.69 & 10.87*&   & 41.60 & B & -2.7 \\
IC1625          &  86.20 & 10.90 &   & 41.75 & B & -3.2 \\
IC1633          &  93.81 & 11.09 &   & 42.79 & N & -3.9 \\
IC1729          &  18.09 &  9.29 & $<$ & 40.00 & B & -4.0 \\
IC1860          &  90.15 & 10.62 &   & 42.71 & B & -4.7 \\
IC2006          &  18.11 &  9.88 & $<$ & 41.03 & B & -4.3 \\
IC2035          &  16.52 &  9.64 & $<$ & 40.62 & B & -2.3 \\
IC2311          &  22.11 &  9.88 & $<$ & 40.22 & B & -4.6 \\
IC2533          &  31.45 & 10.00 & $<$ & 40.44 & B & -3.0 \\
IC2552          &  38.37 & 10.00*& $<$ & 40.69 & B & -3.0 \\
IC2597          &  58.34 & 10.58 & $<$ & 41.03 & B & -3.9 \\
IC3896          &  25.29 &  9.97*& $<$ & 40.50 & B & -4.8 \\
IC3986          &  59.49 & 10.41*&   & 40.30 & N & -4.0 \\
IC4197          &  38.63 &  9.95 & $<$ & 40.73 & B & -3.1 \\
IC4296          &  47.56 & 10.90 &   & 41.53 & N & -4.8 \\
IC4329          &  58.83 & 10.86*& $<$ & 41.82 & F & -3.0 \\
IC4765          &  58.20 & 10.79*&   & 41.83 & N & -3.9 \\
IC4797          &  33.31 & 10.31 & $<$ & 40.99 & B & -3.9 \\
IC4889          &  29.51 & 10.42 & $<$ & 40.80 & B & -4.4 \\
IC4943          &  34.67 &  9.90 & $<$ & 40.64 & B & -4.9 \\
IC5181          &  24.63 &  9.97 & $<$ & 40.28 & B & -2.1 \\
IC5250          &  41.53 & 10.59 &   & 40.14 & N & -2.4 \\
IC5269          &  24.52 &  9.69*& $<$ & 40.46 & R & -1.8 \\
IC5358          & 113.09 & 10.86 &   & 43.58 & B & -3.9 \\
NGC57           &  55.21 & 10.61 &   & 41.65 & B & -4.9 \\
NGC127          &  48.53 &  9.43 & $<$ & 41.16 & F & -2.0 \\
NGC128          &  48.53 & 10.50 & $<$ & 41.15 & F &  5.0 \\
NGC130          &  48.53 &  9.60 & $<$ & 41.18 & F & -3.0 \\
NGC147          &   0.65 &  7.92 & $<$ & 37.45 & B & -4.8 \\
NGC185          &   0.62 &  8.07 & $<$ & 37.36 & B & -4.8 \\
NGC205          &   0.72 &  8.40 & $<$ & 37.61 & B & -4.8 \\
NGC221(M32)     &   0.72 &  8.36 &   & 37.77 & N & -4.7 \\
NGC227          &  71.01 & 10.65 & $<$ & 41.23 & R & -3.6 \\
NGC315          &  58.88 & 11.07 &   & 41.58 & N & -4.0 \\
NGC383          &  56.49 & 10.86 &   & 41.38 & N & -2.9 \\
NGC404          &   0.72 &  7.42 & $<$ & 37.23 & B & -2.8 \\
NGC410          &  56.75 & 10.82 &   & 41.91 & B & -4.3 \\
NGC439          &  74.60 & 10.94*&   & 41.71 & B & -3.2 \\
NGC499          &  55.21 & 10.57 &   & 42.29 & N & -2.8 \\
NGC507          &  67.19 & 10.96 &   & 42.70 & N & -3.2 \\
NGC529          &  65.96 & 10.57 &   & 40.60 & N & -3.0 \\
NGC533          &  63.68 & 10.90 &   & 42.23 & N & -4.8 \\
NGC541          &  63.39 & 10.66 &   & 40.84 & N & -3.8 \\
NGC545          &  63.39 & 10.51 &   & 41.29 & N & -2.9 \\
NGC547          &  63.39 & 10.92 &   & 40.70 & N & -4.7 \\
NGC568          &  73.24 & 10.49 &   & 41.49 & B & -3.0 \\
NGC584          &  22.18 & 10.36 & $<$ & 40.09 & B & -4.6 \\
NGC596          &  22.28 & 10.21 & $<$ & 39.60 & N & -4.3 \\
NGC636          &  22.28 & 10.00 & $<$ & 40.11 & B & -4.8 \\
NGC708          &  55.21 & 10.74 &   & 43.03 & B & -4.8 \\
NGC720          &  20.80 & 10.38 &   & 40.61 & N & -4.8 \\
NGC741          &  61.09 & 10.90 &   & 41.73 & N & -4.8 \\
NGC777          &  55.21 & 10.68 &   & 42.08 & B & -4.8 \\
NGC821          &  20.99 & 10.16 & $<$ & 40.33 & B & -4.8 \\
NGC855          &   9.33 &  8.89 & $<$ & 39.77 & B & -4.8 \\
NGC984          &  59.08 & 10.21 & $<$ & 41.37 & F & -1.3 \\
NGC1016         &  73.79 & 10.95 &   & 41.28 & N & -4.9 \\
NGC1044         &  85.67 & 10.29 & $<$ & 41.17 & F & -3.0 \\
NGC1052         &  17.70 & 10.12 &   & 40.31 & N & -4.7 \\
NGC1167         &  67.67 & 10.50*& $<$ & 41.31 & F & -2.4 \\
NGC1172         &  28.71 & 10.10 & $<$ & 40.59 & B & -3.9 \\
NGC1199         &  28.71 & 10.24 &   & 39.42 & N & -4.7 \\
NGC1201         &  20.67 & 10.16*& $<$ & 40.26 & B & -2.5 \\
NGC1209         &  28.71 & 10.19 & $<$ & 40.62 & B & -4.8 \\
NGC1265         & 102.45 & 10.92 & $<$ & 40.45 & N & -4.0 \\
NGC1316         &  18.11 & 10.93 &   & 40.87 & N & -1.7 \\
NGC1332         &  19.68 & 10.27 &   & 40.53 & N & -2.9 \\
NGC1336         &  18.11 &  9.46 & $<$ & 40.29 & B & -3.0 \\
NGC1339         &  18.11 &  9.73 & $<$ & 40.21 & B & -4.2 \\
NGC1340         &  18.11 & 10.20 & $<$ & 40.27 & B & -3.9 \\
NGC1344         &  18.11 & 10.30 & $<$ & 39.48 & N & -3.9 \\
NGC1351         &  18.11 &  9.78 & $<$ & 40.33 & B & -3.1 \\
NGC1366         &  18.11 &  9.56 & $<$ & 40.32 & B & -2.3 \\
NGC1374         &  18.11 &  9.98 &   & 39.89 & N & -4.5 \\
NGC1375         &  18.11 &  9.58 & $<$ & 38.60 & N & -2.0 \\
NGC1379         &  18.11 & 10.09 &   & 39.24 & N & -4.8 \\
NGC1380         &  18.11 & 10.46 &   & 40.09 & N & -2.3 \\
NGC1380A        &  18.11 &  9.65 & $<$ & 38.96 & N & -1.9 \\
NGC1381         &  18.11 &  9.79 &   & 39.07 & N & -2.0 \\
NGC1387         &  18.11 & 10.03 &   & 40.48 & F & -2.9 \\
NGC1389         &  18.11 &  9.71*& $<$ & 40.08 & F & -2.9 \\
NGC1399         &  18.11 & 10.52 &   & 41.63 & N & -4.5 \\
NGC1395         &  20.51 & 10.44 &   & 40.89 & N & -4.8 \\
NGC1400         &  20.51 & 10.14 &   & 40.12 & N & -3.7 \\
NGC1404         &  18.11 & 10.35 &   & 41.19 & N & -4.7 \\
NGC1407         &  20.61 & 10.58 &   & 41.00 & N & -4.6 \\
NGC1411         &  10.56 &  9.34 & $<$ & 39.63 & B & -3.0 \\
NGC1419         &  18.11 &  9.34 & $<$ & 40.37 & B & -4.8 \\
NGC1426         &  20.61 &  9.92 & $<$ & 40.01 & B & -4.6 \\
NGC1427         &  18.11 & 10.00 &   & 39.85 & N & -4.0 \\
NGC1439         &  20.61 & 10.00 & $<$ & 40.00 & B & -4.7 \\
NGC1497         &  84.12 & 10.41*& $<$ & 41.33 & F & -2.0 \\
NGC1510         &  10.01 &  8.80 & $<$ & 39.76 & F & -2.0 \\
NGC1537         &  16.44 & 10.02 & $<$ & 39.75 & B & -3.3 \\
NGC1549         &  14.45 & 10.28 &   & 39.92 & N & -4.3 \\
NGC1550         &  48.49 & 10.33 &   & 42.80 & B & -3.9 \\
NGC1553         &  14.45 & 10.63 &   & 40.52 & N & -2.3 \\
NGC1573         &  51.52 & 10.72 &   & 41.33 & B & -4.9 \\
NGC1574         &  14.45 & 10.01 & $<$ & 40.32 & F & -2.9 \\
NGC1581         &  14.45 &  9.06 & $<$ & 39.86 & B & -3.0 \\
NGC1587         &  44.87 & 10.51 &   & 40.64 & N & -4.8 \\
NGC1600         &  59.98 & 11.03 &   & 41.54 & B & -4.8 \\
NGC1705         &   4.87 &  8.44*&   & 38.81 & N & -3.0 \\
NGC1947         &  13.43 & 10.18 & $<$ & 40.05 & F & -3.2 \\
NGC2089         &  38.07 & 10.18*& $<$ & 40.56 & B & -3.0 \\
NGC2271         &  32.16 &  9.94*& $<$ & 40.66 & B & -3.2 \\
NGC2272         &   0.72 &  6.81 & $<$ & 37.36 & B & -3.0 \\
NGC2292         &  28.33 & 10.36*& $<$ & 40.60 & B & -2.1 \\
NGC2293         &  23.92 & 10.03 & $<$ & 40.21 & B & -1.1 \\
NGC2300         &  27.67 & 10.41 &   & 41.16 & N & -3.5 \\
NGC2305         &  45.92 & 10.60 &   & 41.67 & B & -4.8 \\
NGC2314         &  48.53 & 10.44 &   & 40.91 & R & -4.7 \\
NGC2325         &  29.79 & 10.60 &   & 40.70 & B & -4.6 \\
NGC2328         &  12.20 &  9.02 & $<$ & 39.56 & B & -2.9 \\
NGC2329         &  71.12 & 10.73 &   & 42.12 & B & -3.0 \\
NGC2340         &  73.79 & 11.04 &   & 42.08 & B & -4.9 \\
NGC2380         &  21.11 &  9.93 & $<$ & 40.20 & B & -2.2 \\
NGC2434         &  14.06 &  9.89 &   & 39.90 & B & -4.8 \\
NGC2444         &  50.82 &  9.92 & $<$ & 41.29 & F & -2.0 \\
NGC2488         & 117.12 & 10.97 &   & 42.56 & B & -3.0 \\
NGC2502         &  11.07 &  9.00 & $<$ & 39.36 & B & -2.1 \\
NGC2562         &  59.43 & 10.18 & $<$ & 41.20 & F & -0.1 \\
NGC2563         &  59.43 & 10.54 &   & 41.63 & N & -2.0 \\
NGC2577         &  28.68 &  9.74 &   & 40.19 & N & -3.0 \\
NGC2629         &  52.13 & 10.29 & $<$ & 40.94 & F & -3.2 \\
NGC2634         &  33.49 & 10.07 & $<$ & 40.46 & B & -4.8 \\
NGC2663         &  27.42 & 10.95 &   & 40.45 & B & -4.6 \\
NGC2685         &  15.85 &  9.80 & $<$ & 40.10 & F & -1.1 \\
NGC2693         &  62.81 & 10.74 & $<$ & 41.29 & F & -4.8 \\
NGC2694         &  62.81 &  9.78 & $<$ & 41.31 & F & -4.9 \\
NGC2695         &  27.42 &  9.97 & $<$ & 40.39 & B & -2.1 \\
NGC2716         &  46.44 & 10.44 & $<$ & 40.67 & F & -1.2 \\
NGC2768         &  20.89 & 10.57 &   & 40.38 & N & -4.4 \\
NGC2778         &  29.24 &  9.80 & $<$ & 40.33 & B & -4.8 \\
NGC2832         &  85.90 & 11.06 &   & 41.62 & N & -4.3 \\
NGC2859         &  23.92 & 10.21 & $<$ & 39.98 & F & -1.2 \\
NGC2865         &  36.48 & 10.48 & $<$ & 40.49 & B & -4.1 \\
NGC2880         &  23.55 &  9.95 & $<$ & 40.06 & B & -2.6 \\
NGC2887         &  35.01 & 10.17 & $<$ & 40.69 & B & -3.2 \\
NGC2888         &  27.12 &  9.62 & $<$ & 40.23 & B & -4.0 \\
NGC2904         &  29.35 &  9.75 & $<$ & 40.37 & B & -3.2 \\
NGC2911         &  41.30 & 10.52 & $<$ & 40.96 & F & -2.1 \\
NGC2974         &  28.31 & 10.50 &   & 40.58 & F & -4.8 \\
NGC2986         &  30.34 & 10.51 &   & 40.96 & B & -4.6 \\
NGC3065         &  30.04 &  9.74 &   & 41.01 & F & -2.0 \\
NGC3073         &  18.49 &  9.09 & $<$ & 39.77 & B & -2.8 \\
NGC3078         &  33.42 & 10.48 &   & 40.72 & B & -4.8 \\
NGC3087         &  34.67 & 10.52 & $<$ & 40.56 & B & -4.2 \\
NGC3091         &  50.82 & 10.75 &   & 41.63 & N & -4.5 \\
NGC3115         &   8.83 & 10.10 &   & 39.72 & N & -2.8 \\
NGC3136         &  19.11 & 10.07 & $<$ & 40.26 & B & -4.8 \\
NGC3156         &  19.95 &  9.71 & $<$ & 40.00 & B & -2.5 \\
NGC3158         &  86.30 & 10.96 &   & 41.71 & B & -4.8 \\
NGC3193         &  21.58 & 10.15 &   & 39.96 & N & -4.7 \\
NGC3222         &  75.09 & 10.46 & $<$ & 41.33 & F & -2.1 \\
NGC3224         &  38.55 & 10.16 & $<$ & 40.65 & B & -3.7 \\
NGC3226         &  21.58 & 10.12 &   & 40.20 & N & -4.8 \\
NGC3250         &  37.67 & 10.71 & $<$ & 40.65 & B & -4.8 \\
NGC3258         &  38.37 & 10.48 &   & 41.17 & F & -4.3 \\
NGC3268         &  38.37 & 10.48 &   & 40.53 & N & -4.3 \\
NGC3271         &  47.81 & 10.43*&   & 41.06 & N & -1.8 \\
NGC3311         &  58.34 & 10.76 &   & 42.19 & N & -3.4 \\
NGC3375         &  30.92 &  9.80 & $<$ & 40.43 & B & -2.0 \\
NGC3377         &  10.00 &  9.72 & $<$ & 39.60 & B & -4.8 \\
NGC3379         &  10.00 & 10.06 & $<$ & 39.54 & B & -4.8 \\
NGC3384         &  10.00 &  9.85 & $<$ & 39.52 & B & -2.6 \\
NGC3458         &  27.03 &  9.77*& $<$ & 40.61 & F & -2.0 \\
NGC3516         &  38.37 & 10.36 &   & 43.11 & N & -2.0 \\
NGC3557         &  32.21 & 10.76 &   & 40.58 & N & -4.8 \\
NGC3585         &  16.07 & 10.39 &   & 39.79 & N & -4.5 \\
NGC3599         &  19.77 &  9.66 & $<$ & 39.23 & N & -2.0 \\
NGC3605         &  19.77 &  9.47 &   & 39.08 & N & -4.7 \\
NGC3606         &  37.75 &  9.98*& $<$ & 40.69 & B & -4.9 \\
NGC3607         &  19.77 & 10.46 &   & 40.54 & F & -3.1 \\
NGC3608         &  19.77 & 10.11 &   & 40.01 & N & -4.8 \\
NGC3610         &  27.29 & 10.40 &   & 39.83 & N & -4.2 \\
NGC3613         &  27.29 & 10.36 & $<$ & 40.12 & B & -4.7 \\
NGC3617         &  27.39 &  9.59 & $<$ & 40.40 & B & -3.9 \\
NGC3640         &  22.91 & 10.43 &   & 39.92 & N & -4.8 \\
NGC3656         &  41.00 & 10.10 & $<$ & 40.61 & B &  0.0 \\
NGC3658         &  30.76 &  9.92 & $<$ & 39.77 & N & -2.2 \\
NGC3665         &  30.76 & 10.70 &   & 40.60 & N & -2.1 \\
NGC3706         &  37.21 & 10.38 & $<$ & 41.19 & B & -3.3 \\
NGC3818         &  21.48 &  9.80 & $<$ & 40.16 & B & -4.8 \\
NGC3842         &  82.04 & 10.92 &   & 41.80 & N & -4.9 \\
NGC3862         &  82.04 & 10.56 &   & 41.90 & B & -4.9 \\
NGC3894         &  46.37 & 10.47 &   & 41.19 & F & -4.1 \\
NGC3904         &  17.86 & 10.06 & $<$ & 40.74 & B & -4.6 \\
NGC3923         &  17.86 & 10.52 &   & 40.66 & N & -4.5 \\
NGC3962         &  21.68 & 10.28 & $<$ & 40.22 & B & -4.8 \\
NGC3990         &  12.16 &  8.99 &   & 38.60 & N & -2.7 \\
NGC3998         &  17.46 & 10.08 &   & 41.51 & N & -2.1 \\
NGC4024         &  20.84 &  9.77 & $<$ & 40.04 & B & -3.2 \\
NGC4033         &  19.25 &  9.71 & $<$ & 40.00 & B & -4.5 \\
NGC4036         &  21.73 & 10.24 & $<$ & 40.03 & B & -2.5 \\
NGC4073         &  79.43 & 11.07 &   & 42.38 & N & -4.1 \\
NGC4104         & 111.87 & 11.05*&   & 42.66 & N & -2.0 \\
NGC4105         &  22.85 & 10.25*&   & 40.42 & F & -4.6 \\
NGC4125         &  25.94 & 10.80 &   & 40.94 & N & -4.8 \\
NGC4168         &  33.73 & 10.40 &   & 40.56 & F & -4.8 \\
NGC4203         &  16.22 &  9.89 &   & 41.18 & N & -2.7 \\
NGC4215         &  31.48 &  9.98 & $<$ & 40.46 & F & -0.8 \\
NGC4233         &  31.48 & 10.02 & $<$ & 39.22 & N & -2.0 \\
NGC4239         &  16.75 &  9.24 & $<$ & 39.85 & B & -4.7 \\
NGC4251         &  16.22 & 10.02 & $<$ & 39.65 & F & -1.9 \\
NGC4261         &  31.48 & 10.70 &   & 41.21 & N & -4.8 \\
NGC4262         &  15.92 &  9.65 & $<$ & 39.82 & B & -2.7 \\
NGC4267         &  15.92 &  9.88 & $<$ & 39.90 & B & -2.7 \\
NGC4278         &  16.22 & 10.24 &   & 40.36 & N & -4.8 \\
NGC4283         &  16.22 &  9.46 & $<$ & 39.22 & N & -4.8 \\
NGC4291         &  24.55 & 10.00 &   & 40.89 & N & -4.8 \\
NGC4339         &  15.92 &  9.71 & $<$ & 39.86 & B & -4.7 \\
NGC4340         &  15.92 &  9.84 &   & 39.75 & R & -1.2 \\
NGC4350         &  15.92 &  9.85 & $<$ & 39.77 & F & -1.8 \\
NGC4365         &  15.92 & 10.34 &   & 40.25 & N & -4.8 \\
NGC4374(M84)    &  15.92 & 10.57 &   & 40.83 & N & -4.0 \\
NGC4382         &  15.92 & 10.64 &   & 40.33 & F & -1.3 \\
NGC4386         &  24.55 &  9.90 & $<$ & 39.93 & F & -2.1 \\
NGC4387         &  15.92 &  9.47 &   & 39.71 & N & -4.8 \\
NGC4406(M86)    &  15.92 & 10.66 &   & 42.05 & N & -4.7 \\
NGC4417         &  15.92 &  9.77 & $<$ & 40.68 & F & -1.9 \\
NGC4425         &  15.92 &  9.64 & $<$ & 39.86 & F & -0.7 \\
NGC4434         &  15.92 &  9.45 & $<$ & 39.83 & B & -4.8 \\
NGC4435         &  15.92 & 10.01 & $<$ & 40.13 & F & -2.1 \\
NGC4458         &  16.14 &  9.51 &   & 39.84 & F & -4.8 \\
NGC4459         &  15.92 & 10.20 &   & 40.17 & R & -1.4 \\
NGC4464         &  15.92 &  9.20 & $<$ & 39.81 & B &  1.7 \\
NGC4467         &  15.92 &  8.59 & $<$ & 39.29 & F & -4.9 \\
NGC4472(M49)    &  15.92 & 10.90 &   & 41.43 & N & -4.7 \\
NGC4473         &  16.14 & 10.15 &   & 40.14 & F & -4.8 \\
NGC4474         &  15.92 &  9.64 & $<$ & 39.85 & F & -2.0 \\
NGC4476         &  15.92 &  9.43 & $<$ & 40.27 & R & -3.0 \\
NGC4477         &  15.92 & 10.13 &   & 40.26 & N & -1.9 \\
NGC4478         &  15.92 &  9.79 & $<$ & 40.41 & R & -4.8 \\
NGC4479         &  15.92 &  9.23 & $<$ & 39.70 & F & -1.9 \\
NGC4486(M87)    &  15.92 & 10.85 &   & 42.95 & B & -4.3 \\
NGC4489         &  15.92 &  9.46 & $<$ & 39.84 & B & -4.8 \\
NGC4494         &  21.28 & 10.62 & $<$ & 40.10 & B & -4.8 \\
NGC4503         &  15.92 &  9.77 & $<$ & 39.88 & B & -2.0 \\
NGC4507         &  45.24 & 10.33 & $<$ & 41.40 & F &  1.9 \\
NGC4515         &  15.92 &  9.24 & $<$ & 39.80 & B & -3.0 \\
NGC4526         &  15.92 & 10.47 &   & 39.87 & N & -1.9 \\
NGC4550         &  15.92 &  9.72 &   & 39.78 & N & -2.1 \\
NGC4551         &  15.92 &  9.58 & $<$ & 39.09 & N & -4.8 \\
NGC4552(M89)    &  15.92 & 10.29 &   & 40.71 & N & -4.6 \\
NGC4555         &  90.33 & 10.86*&   & 41.85 & N & -4.8 \\
NGC4564         &  15.92 &  9.86 & $<$ & 39.85 & B & -4.7 \\
NGC4578         &  15.92 &  9.78 & $<$ & 39.99 & F & -2.0 \\
NGC4581         &  15.92 &  9.26 & $<$ & 39.96 & B & -4.4 \\
NGC4589         &  24.55 & 10.33 &   & 40.36 & R & -4.8 \\
NGC4621         &  15.92 & 10.32 &   & 40.02 & R & -4.8 \\
NGC4627         &  12.13 &  9.13 &   & 39.92 & B & -4.7 \\
NGC4636         &  15.92 & 10.51 &   & 41.59 & N & -4.8 \\
NGC4638         &  15.92 &  9.80 &   & 39.59 & N & -2.7 \\
NGC4645         &  32.03 & 10.09*& $<$ & 40.57 & B & -3.9 \\
NGC4648         &  24.55 &  9.87 & $<$ & 39.89 & B & -4.9 \\
NGC4649(M60)    &  15.92 & 10.73 &   & 41.28 & N & -4.6 \\
NGC4660         &  15.92 &  9.74 & $<$ & 39.39 & N & -4.7 \\
NGC4697         &  15.14 & 10.55 &   & 40.12 & N & -4.8 \\
NGC4696         &  37.01 & 10.99*&   & 43.23 & N & -3.9 \\
NGC4709         &  59.31 & 10.94 &   & 41.00 & N & -4.5 \\
NGC4733         &  15.92 &  9.51 & $<$ & 39.73 & B & -3.9 \\
NGC4742         &  12.42 &  9.56 & $<$ & 39.80 & B & -4.8 \\
NGC4751         &  23.97 &  9.76*& $<$ & 40.31 & B & -2.9 \\
NGC4753         &  20.23 & 10.46 &   & 39.99 & F & -1.6 \\
NGC4754         &  15.92 & 10.00 & $<$ & 39.73 & B & -2.5 \\
NGC4756         &  53.93 & 10.30 &   & 41.72 & F & -2.9 \\
NGC4760         &  63.39 & 11.02 &   & 41.58 & B & -4.8 \\
NGC4762         &  15.92 & 10.16 &   & 40.13 & F & -1.8 \\
NGC4767         &  37.39 & 10.33 & $<$ & 40.71 & B & -4.0 \\
NGC4782         &  63.39 & 11.37 &   & 41.61 & F & -4.8 \\
NGC4880         &  20.28 &  9.83 & $<$ & 40.20 & F & -1.5 \\
NGC4839         &  87.90 & 10.90 &   & 40.45 & N & -4.0 \\
NGC4889         &  88.31 & 11.19 &   & 42.76 & N & -4.3 \\
NGC4915         &  43.85 & 10.28 & $<$ & 40.87 & B & -4.7 \\
NGC4936         &  41.07 & 10.71*&   & 41.69 & B & -4.6 \\
NGC4946         &  38.53 &  9.98 & $<$ & 40.79 & B & -4.1 \\
NGC4976         &  11.43 &  9.97 & $<$ & 39.73 & B & -4.4 \\
NGC4993         &  37.49 & 10.01*& $<$ & 40.71 & B & -3.0 \\
NGC5011         &  38.76 & 10.40 & $<$ & 40.82 & B & -4.8 \\
NGC5018         &  30.20 & 10.57 & $<$ & 40.53 & B & -4.5 \\
NGC5044         &  30.20 & 10.70 &   & 42.74 & N & -4.8 \\
NGC5061         &  18.28 & 10.28 &   & 39.68 & N & -4.3 \\
NGC5077         &  30.20 & 10.26 &   & 40.48 & F & -4.8 \\
NGC5084         &  16.90 & 10.18 &   & 40.49 & F & -1.6 \\
NGC5087         &  18.71 & 10.03 &   & 40.36 & B & -3.0 \\
NGC5090         &  42.23 & 10.41 &   & 41.49 & B & -4.9 \\
NGC5102         &   4.16 &  9.29 &   & 38.03 & N & -3.0 \\
NGC5128         &   3.89 & 10.45 &   & 40.10 & N & -2.3 \\
NGC5129         &  91.20 & 10.91 &   & 42.14 & N & -4.9 \\
NGC5153         &  55.65 & 10.55 &   & 40.50 & N & -4.8 \\
NGC5173         &  34.99 & 10.04 & $<$ & 40.36 & B & -4.9 \\
NGC5193         &  47.41 & 10.55 &   & 40.58 & N & -4.2 \\
NGC5195         &   9.12 &  9.93 &   & 39.42 & F &  0.1 \\
NGC5198         &  34.99 & 10.28 & $<$ & 40.38 & B & -4.8 \\
NGC5216         &  42.33 & 10.02*&   & 40.85 & B & -4.9 \\
NGC5253         &   3.64 &  8.96 & $<$ & 38.77 & B &  7.7 \\
NGC5273         &  17.09 &  9.68 &   & 39.86 & N & -1.9 \\
NGC5306         &  96.41 & 10.91 &   & 41.50 & N & -2.1 \\
NGC5308         &  27.80 & 10.20 & $<$ & 40.01 & B & -2.0 \\
NGC5318         &  59.49 & 10.24 & $<$ & 41.04 & F & -2.0 \\
NGC5322         &  27.80 & 10.67 &   & 40.21 & N & -4.8 \\
NGC5328         &  61.40 & 10.70 &   & 41.88 & B & -4.8 \\
NGC5353         &  34.67 & 10.56 &   & 41.00 & F & -2.1 \\
NGC5354         &  33.32 & 10.30 & $<$ & 40.84 & F & -2.0 \\
NGC5363         &  15.79 & 10.17 &   & 40.14 & F &  0.0 \\
NGC5382         &  58.19 & 10.32 &   & 40.14 & N & -2.0 \\
NGC5419         &  53.44 & 10.88 &   & 41.80 & N & -4.4 \\
NGC5473         &  28.18 & 10.21 & $<$ & 40.09 & B & -2.7 \\
NGC5485         &  28.18 & 10.25 & $<$ & 40.12 & B & -2.0 \\
NGC5507         &  25.85 &  9.63 &   & 39.75 & N & -2.1 \\
NGC5532         &  98.13 & 11.02 &   & 41.63 & F & -2.0 \\
NGC5546         &  98.53 & 10.86 &   & 42.02 & B & -4.9 \\
NGC5574         &  21.68 &  9.57 & $<$ & 40.14 & B & -2.8 \\
NGC5576         &  21.68 & 10.16 & $<$ & 40.14 & B & -4.8 \\
NGC5582         &  18.40 &  9.73 & $<$ & 39.82 & B & -4.9 \\
NGC5638         &  21.68 & 10.09 & $<$ & 40.20 & B & -4.8 \\
NGC5687         &  31.49 & 10.15 & $<$ & 40.14 & B & -3.0 \\
NGC5812         &  24.55 & 10.19 & $<$ & 40.32 & B & -4.8 \\
NGC5831         &  22.91 & 10.03 & $<$ & 40.25 & B & -4.8 \\
NGC5838         &  22.91 & 10.20 &   & 40.02 & F & -2.7 \\
NGC5845         &  22.91 &  9.56 & $<$ & 39.94 & N & -4.9 \\
NGC5846         &  22.91 & 10.66 &   & 41.65 & N & -4.7 \\
NGC5866         &  13.18 & 10.32 &   & 39.69 & F & -1.3 \\
NGC5898         &  23.88 & 10.22 & $<$ & 40.31 & B & -4.2 \\
NGC5903         &  23.88 & 10.28 & $<$ & 40.33 & B & -4.6 \\
NGC5982         &  37.50 & 10.53 &   & 41.16 & B & -4.8 \\
NGC6027         &  60.88 &  9.63 & $<$ & 41.39 & F & -1.4 \\
NGC6034         & 137.28 & 10.63 &   & 42.19 & F & -4.0 \\
NGC6127         &  65.01 & 10.61 &   & 41.40 & B & -4.9 \\
NGC6137         & 112.72 & 11.13 &   & 42.14 & B & -4.8 \\
NGC6146         & 107.65 & 10.92 & $<$ & 41.90 & F & -4.7 \\
NGC6160         & 127.65 & 10.72*&   & 42.45 & B & -4.9 \\
NGC6166         & 108.64 & 11.20 &   & 43.93 & B & -4.3 \\
NGC6173         & 107.65 & 11.09 &   & 42.15 & B & -4.9 \\
NGC6269         & 139.68 & 11.15 &   & 42.86 & B & -4.8 \\
NGC6305         &  33.72 &  9.95 & $<$ & 40.94 & B & -3.0 \\
NGC6407         &  58.67 & 10.58*&   & 41.94 & B & -2.0 \\
NGC6482         &  54.69 & 10.72 &   & 42.08 & N & -4.8 \\
NGC6487         & 105.39 & 11.07 &   & 41.70 & B & -4.9 \\
NGC6673         &  12.29 &  9.34 & $<$ & 40.06 & B & -3.9 \\
NGC6684         &   8.47 &  9.73 &   & 38.93 & N & -1.8 \\
NGC6703         &  29.92 & 10.37 & $<$ & 40.03 & B & -2.8 \\
NGC6776         &  70.41 & 10.66 &   & 40.79 & N & -4.1 \\
NGC6841         &   0.15 &  5.01 & $<$ & 36.06 & B & -3.9 \\
NGC6851         &  34.67 & 10.30 & $<$ & 40.64 & B & -4.8 \\
NGC6861         &  34.67 & 10.42 & $<$ & 40.65 & B & -2.7 \\
NGC6868         &  34.67 & 10.58*&   & 41.23 & N & -4.4 \\
NGC6876         &  48.56 & 10.83*&   & 41.51 & F & -4.9 \\
NGC6880         &  50.77 & 10.33 & $<$ & 41.10 & F & -1.0 \\
NGC6909         &  34.67 & 10.27 & $<$ & 40.78 & B & -4.1 \\
NGC6920         &  34.13 &  9.83 & $<$ & 40.95 & B & -2.0 \\
NGC6958         &  34.79 & 10.34 & $<$ & 40.68 & B & -3.5 \\
NGC6963         &  58.93 &  9.73 & $<$ & 40.97 & F & -2.3 \\
NGC6964         &  51.83 & 10.02 & $<$ & 40.84 & F & -4.4 \\
NGC7007         &  37.39 & 10.16 & $<$ & 40.70 & B & -2.9 \\
NGC7029         &  34.97 & 10.34 & $<$ & 40.57 & B & -4.4 \\
NGC7041         &  23.39 & 10.09 & $<$ & 40.24 & B & -3.0 \\
NGC7049         &  27.27 & 10.37 &   & 41.01 & B & -2.1 \\
NGC7097         &  29.24 & 10.13 &   & 40.28 & N & -4.8 \\
NGC7144         &  21.38 & 10.14 &   & 39.64 & N & -4.8 \\
NGC7145         &  21.38 & 10.04 & $<$ & 40.25 & B & -4.8 \\
NGC7166         &  30.43 & 10.02 & $<$ & 40.45 & B & -2.9 \\
NGC7168         &  34.57 & 10.12 & $<$ & 40.59 & B & -4.7 \\
NGC7173         &  32.65 & 10.04 &   & 40.86 & N & -4.1 \\
NGC7176         &  32.39 & 10.27 &   & 40.80 & N & -4.6 \\
NGC7180         &  16.05 &  9.18 & $<$ & 40.07 & B & -2.4 \\
NGC7185         &  24.04 &  9.59 & $<$ & 40.38 & B & -3.0 \\
NGC7192         &  35.77 & 10.42 &   & 40.85 & B & -3.9 \\
NGC7196         &  36.51 & 10.35 &   & 40.95 & B & -4.8 \\
NGC7236         & 105.60 & 10.39 & $<$ & 41.63 & F & -3.0 \\
NGC7237         & 105.31 & 10.29 & $<$ & 41.76 & F & -3.0 \\
NGC7252         &  52.48 & 10.66 &   & 40.50 & N & -2.0 \\
NGC7265         &  68.17 & 10.56 &   & 41.70 & B & -2.7 \\
NGC7332         &  15.28 &  9.86 & $<$ & 40.01 & F & -1.9 \\
NGC7385         & 105.75 & 11.11 &   & 41.79 & B & -4.8 \\
NGC7454         &  24.32 &  9.95 & $<$ & 40.20 & B & -4.8 \\
NGC7457         &  10.67 &  9.78 & $<$ & 39.49 & B & -2.7 \\
NGC7465         &  24.32 &  9.64 &   & 41.36 & N & -1.9 \\
NGC7484         &  34.69 & 10.16*& $<$ & 40.93 & B & -4.8 \\
NGC7507         &  17.78 & 10.23 & $<$ & 40.77 & B & -4.8 \\
NGC7550         &  69.64 & 10.61 & $<$ & 40.31 & N & -3.0 \\
NGC7562         &  39.99 & 10.46 & $<$ & 40.93 & F & -4.8 \\
NGC7619         &  39.99 & 10.58 &   & 41.63 & N & -4.8 \\
NGC7626         &  39.99 & 10.61 &   & 41.06 & N & -4.8 \\
NGC7768         &  92.04 & 10.92 &   & 41.74 & B & -4.9 \\
NGC7796         &  39.45 & 10.48 &   & 40.74 & N & -3.9 \\
UGC34(Maff I)   &  82.45 & 10.39 & $<$ & 41.29 & B &  2.0 \\
UGC1308         &  55.21 & 10.16*&   & 40.98 & N & -4.9 \\
UGC4956         &  67.63 & 10.45 &   & 41.60 & B & -4.9 \\
UGC5470(Leo I)  &   2.33 &  8.60 & $<$ & 38.52 & B & -4.8 \\
UGC6253(Leo II) &   2.17 &  7.94 & $<$ & 38.51 & B & -4.8 \\
\hline
\caption{\label{Lxtab2} \label{lastpage}Combined catalogue of X--ray luminosities. The
  catalogue contains 401 early--type galaxies and 24 late--type objects which 
  were included in previous catalogues. L$_B$ values are based on B$_T$
  magnitudes, except those marked *, which are based on m$_B$ magnitudes
  (see Section~\ref{catconv}). L$_X$ values are bolometric and T--type
  is taken from LEDA. The source of each L$_X$ value is shown, B signifying 
  Beuing \etal (1999), F = Fabbiano, Kim \& Trinchieri (1992), R = Roberts
  \etal (1991) and N = new values calculated by the authors as described in 
  Section~\ref{Specfit}.}
\end{longtable}
\twocolumn
\label{lastpage}

\end{document}